\documentclass[11pt]{article}
\usepackage{latexsym}
\usepackage{amssymb}
\usepackage{amsmath,amsfonts}
\usepackage{latexsym}
\usepackage{graphicx} 

\catcode`\@=11
\def\numberbysection{\@addtoreset{equation}{section}
        \def\theequation{\thesection.\arabic{equation}}}

\def\be{\begin{equation}} 
\def\ee{\end{equation}} 
\def\ba{\begin{eqnarray}} 
\def\ea{\end{eqnarray}} 
 
\def\ov{\overline} 
\def\I{{\rm Im}} 
\def\R{{\rm Re}} 
\def\Z{\mathbb{Z}}

\def\nl{\nonumber \\}

\def\wt{\widetilde} 
\def\wh{\widehat} 
\def\Tr{{\rm Tr}}

\def\a{\alpha} 
\def\b{\beta} 
 
\def\G{\Gamma} 
\def\D{\Delta} 
\def\d{\delta}

\def\z{\zeta}

\def\l{\lambda}

\def\x{\xi} 
 
\def\p{\pi}

\def\s{\sigma} 
 
\def\t{\tau} 
\def\f{\phi}

\def\c{\chi}

\def\Th{\Theta} 
\def\th{\theta}

\def\u1{\widehat{U(1)}}
\def\su2{\widehat{SU(2)}_1}

\numberbysection

\begin{document}
\begin{titlepage}
\begin{center}

\vskip .6 in
{\LARGE Partition Functions and Stability Criteria}

{\LARGE of Topological Insulators}

\vskip 0.2in
Andrea CAPPELLI \\
{\em INFN \\ Via G. Sansone 1, 50019 Sesto Fiorentino - Firenze, Italy}

Enrico RANDELLINI \\
{\em Dipartimento di Fisica \\ 
Via G. Sansone 1, 50019 Sesto Fiorentino - Firenze, Italy}
\end{center}
\vskip .2 in
\begin{abstract}
The non-chiral edge excitations of quantum spin Hall systems and topological
insulators are described by means of their partition function. The stability
of topological phases protected by time-reversal symmetry is rediscussed in
this context and put in relation with the existence of discrete anomalies and
the lack of modular invariance of the partition function.  The $\Z_2$
characterization of stable topological insulators is extended to systems with
interacting and non-Abelian edge excitations.
\end{abstract}

\vfill

\end{titlepage}
\pagenumbering{arabic}


\section{Introduction}
The study of topological phases of matter has considerably grown in
recent years and new systems have been investigated both theoretically
and experimentally \cite{qz-rev}. 
It is now apparent that some remarkable
topological features of quantum Hall states can occur in a wider
set of systems and thus be more universal and robust. In particular,
non-chiral topological states, such as those of the quantum spin Hall effect,
of topological insulators and of topological superconductors, do not
require strong magnetic fields and exist in three space dimensions
\cite{molen}\cite{3d-ti}.

A characteristic feature of topological states is the existence of
massless edge excitations that are well accounted for by low-energy
effective field theory descriptions \cite{wen-book}.  
The response of topological
states to external disturbances does not occur in the gapped
bulk, but manifests itself through the edge dynamics.  While the
edge excitations of chiral states, such as quantum Hall states
and Chern insulators, are absolutely stable, those of non-chiral
topological states can interact and became gapful, leading to
the decay into topologically trivial phases. 
In some cases, edge interactions are forbidden by  the presence of
(discrete) symmetries: we then speak of
symmetry protected topological phases of matter \cite{wen}.

Topological insulators protected by time-reversal (TR) symmetry have been
first analyzed in free fermion systems using band theory
\cite{kane}\cite{tbt}.  
These systems were found to be characterized by a topological bulk quantity
equal to the $\Z_2$ index $(-1)^N$, where $N$ is the number of fermion edge
modes of each chirality. The odd (even) $N$ cases were shown to be stable
(unstable) upon using a remarkable flux argument by Fu, Kane and
Mele \cite{kane}.  The stability analysis was
then reformulated in terms of the dynamics of edge excitations in
Refs. \cite{ls}\cite{chamon}\cite{vish}, 
by using conformal field theory (CFT) methods \cite{cft}.  
It was shown
that the $\Z_2$ classification extends to interacting systems that are
described by Abelian edge excitations within the so-called $K$-matrix
formalism \cite{wen-book}.

The analysis of Levin and Stern led to the following simple and general 
result for the $\Z_2$ index \cite{ls}:
\be
(-1)^{2\Delta S}, \qquad 2\Delta S=\frac{\sigma_{sH}}{e^*},
\label{ls-index}
\ee 
where $\sigma_{sH}$ is the spin Hall conductance and $e^*$ is the minimal
fractional charge for one of the spin (chiral) components, in units of
$e/2\pi$ and $e$, respectively.  
Their ratio measures the spin $\D S$ of an excitation created
at the edge, as explained later. An odd (even) ratio
corresponds to stable (unstable) systems, as this quantity reduces to the
number of fermion modes in the non-interacting case.

In this paper, we rederive and further extend the $\Z_2$ stability
criterion (\ref{ls-index}) by studying the partition function of edge
excitations. In earlier work, we obtained the general form of the
partition function for both Abelian and non-Abelian quantum Hall
states \cite{cz-mod}\cite{cgt1}\cite{cv}. 
Upon generalizing this analysis to the quantum spin Hall effect
and topological insulators, we can apply the flux argument for
stability to any type of interacting topological insulator and prove
the general validity of (\ref{ls-index}).

Our analysis clarifies that the stability is associated to the
presence of an anomaly in the $Z_2$ symmetry of fermion number parity
at the edge, also equal to the edge spin parity $(-1)^{2S}$.
Actually, the $U(1)_S$ spin symmetry of the quantum spin Hall effect
is explicitly broken to $(-1)^{2S}$ by spin-orbit interaction and
other TR invariant relativistic corrections that are present in generic
topological insulators \cite{qz-rev}. 
The $Z_2$ anomaly is thus the remnant of the
$U(1)_S$ anomaly of the spin Hall effect, which is analogous 
to the $U(1)_Q$ charge anomaly in the Hall effect \cite{wen-book}.

The partition function  is obtained in the double periodic
geometry of the torus made by one circular edge  and
compact Euclidean time for temperature \cite{cft}\cite{cz-mod}.  
This function contains anyonic sectors for each
chirality, that involve sums over charged and neutral edge modes.
This  structure allows one to disentangle the Abelian charged
mode that determines the $\Z_2$ anomaly, from the neutral modes,
either Abelian or non-Abelian, that are transparent. We can thus analyze
the stability (anomaly) for general interacting system.

The response of the topological state, i.e. of its edge dynamics, to
external perturbations can be easily described by using the
partition function. One can see how its different sectors 
 transforms among themselves by inserting fluxes, i.e. under the
effect of an electromagnetic background. Furthermore, the behaviour
under modular transformations, the discrete coordinate changes
respecting the periodicities of the torus, yield the
response to (some type of) gravitational backgrounds: they amount, e.g. to
adding momentum to the system or to swapping
space and time (respectively, the $T$ and $S$ modular transformations).

We find that the stability (instability) of non-chiral edge states is
associated to the impossibility (possibility) of having a modular
invariant partition function that is consistent with time reversal symmetry.
The modular non-invariance of stable topological states is actually a
kind of gravitational anomaly accompanying the spin parity anomaly. In
some systems, we find that flux insertions and modular
transformations act in similar way on the four ``spin sectors'' always present
in fermionic systems, corresponding to periodic and antiperiodic
boundary conditions in space and time \cite{cft}\cite{gso}.  
In other systems, the two backgrounds have different effects
owing to the different couplings of charged and neutral modes.

Modular non-invariance and stability of topological
states have already been discussed in a paper by Ryu and Zhang \cite{rz}
that actually motivated our work. They considered the topological
superconductors, whose neutral Majorana edge modes cannot be analyzed by flux
arguments, and argued that modular non-invariance of the partition function
could be used as a criterion of stability.
Then they showed that the system of $N_f$ Majorana fermions is unstable 
for $N_f=0$ modulo $8$, finding agreement with the study of possible
interactions \cite{fk}\cite{qi}.  Our analysis reproduces this
result, but cannot presently provide a general approach to interacting
topological superconductors. This open issue is discussed in the conclusions.

The paper is organized as follows. In Section two, we briefly 
describe the setting of the problem and recall
the flux insertion argument by Fu, Kane and Mele.
In Section three, we introduce the partition function of topological
insulators with one edge mode per spin, described by
the Luttinger liquid CFT with
central charge $c=1$. We discuss the effect of flux insertions and
modular transformations, and rederive the $\Z_2$ anomaly and the stability
argument in this context.  In Section four, we extend the
analysis to general non-chiral edge CFTs and discuss some
interesting examples.  In Section five, we present our conclusions.
The Appendix contains the details on modular transformations of Abelian
and non-Abelian theories.


\section{Flux argument and stability analysis}

\subsection{Laughlin flux insertion}

\begin{figure}[t]
\begin{center}
\includegraphics[width=7cm]{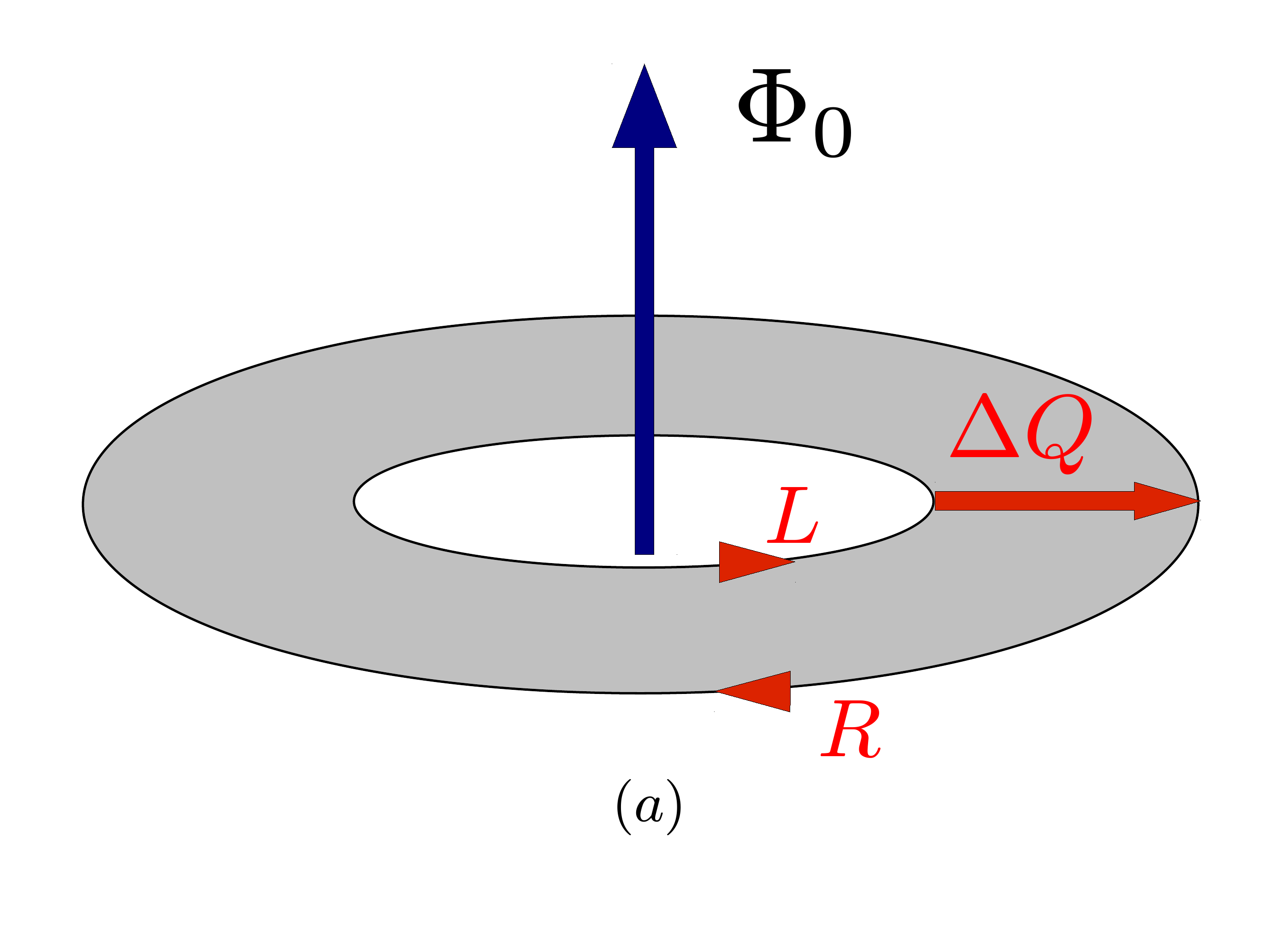}
\hspace{.3cm}
\includegraphics[width=7cm]{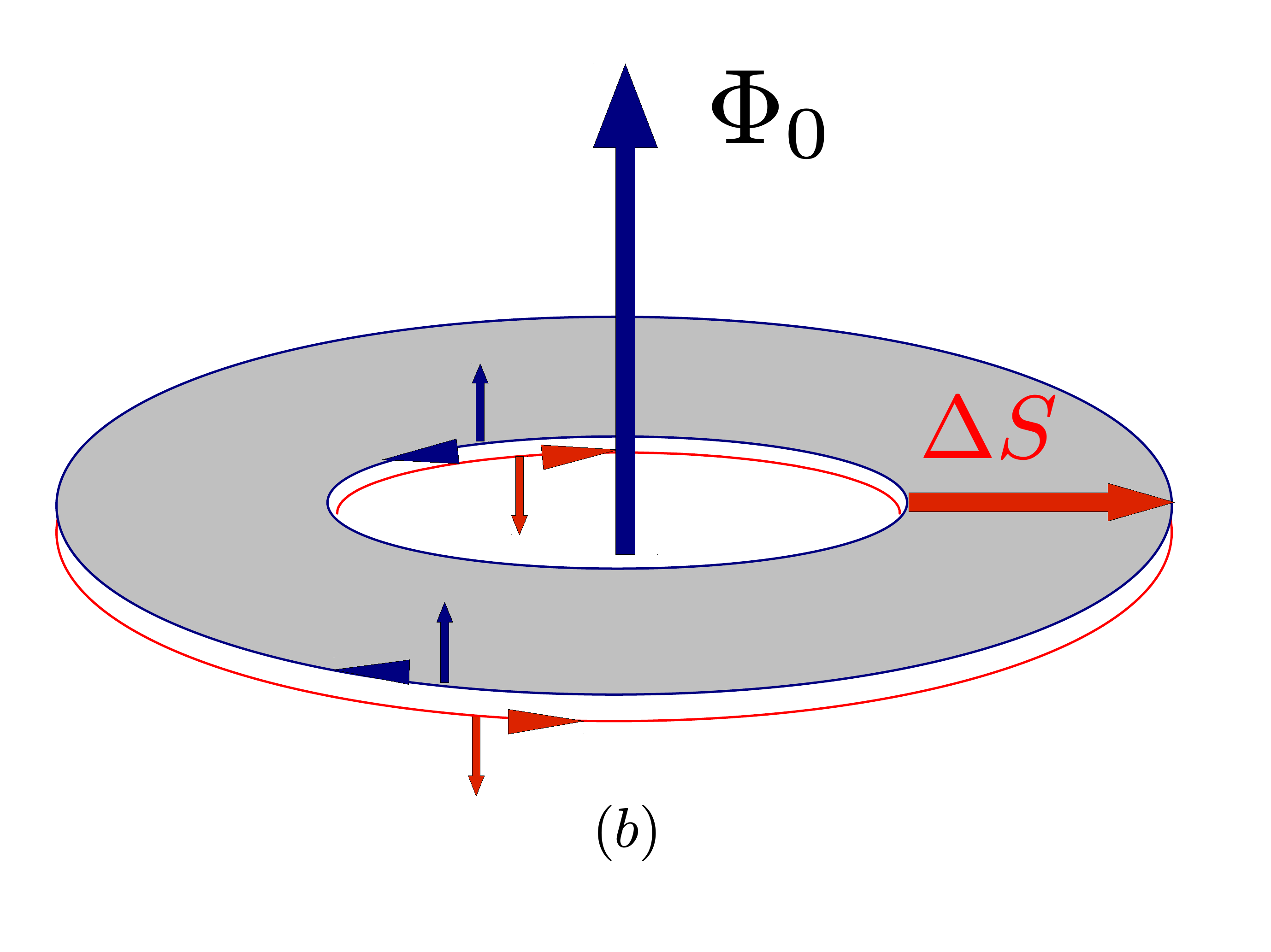}
\caption{Flux insertion in the (a) QHE and (b) QSHE.}
\label{fig1}
\end{center}
\end{figure}

We start by recalling the Laughlin argument for the quantization of
the Hall current in the annulus geometry (see Fig. \ref{fig1}(a)) 
\cite{laugh}.  Upon the adiabatic
insertion of one quantum of flux $\Phi_0$, the bulk Hamiltonian returns to
itself while states in the spectrum drift one into another, leading
to the so-called spectral flow. This amounts to the transport of a
charge equal to the Hall conductivity between the two edges.  
\ba
&&\Phi \to \Phi+ \Phi_0, \qquad\quad
H\left[\Phi+\Phi_0\right]=H\left[\Phi\right],
\nl 
&&Q\to Q+\Delta Q= \nu\ .
\label{nu-hall}
\ea  
From the point of view of the conformal field theory describing one
edge of the annulus, the spectral flow corresponds to the non-conservation
of charge, namely a $U(1)_Q$ chiral anomaly in two dimensions 
\cite{wen-book}\cite{cdtz}.
The associated index theorem for the integrated anomaly reads:
\be
\Delta Q= \int_\infty^\infty\! dt\int_0^{2\pi R}\! dx\ \partial_t J^0_R =
\frac{e\nu}{2\pi} \int\! F = \nu\, e\, n.
\qquad\quad n\in \Z.
\label{int-anomaly}
\ee 
The last term in this expression is a topological quantity, the
first Chern class of the two-dimensional electromagnetic field,
$F=\frac{1}{2}F_{ij}dx^i\wedge dx^j$, whose integer values count the number of
inserted flux quanta. We also recall that the chiral anomaly is an
universal effect that is exact for any strength of the interaction
\cite{cft}.
 
We can interpreted the chiral anomaly of the edge as the response of the
topological bulk to an electromagnetic background, as well
explained in the associated description in terms of the effective hydrodynamic
Chern-Simons theory \cite{wen-book}.  
Other chiral topological phases are also associated to
anomalies, leading to the breaking of continuous symmetries and to
anomalous currents.  These correspond to non-vanishing values of transport
coefficients, such as the Hall conductivity $\sigma_H$ and/or thermal
transport coefficient $ \kappa_H$.  The general analysis of topological
phases characterized by electromagnetic,
gravitational and mixed chiral anomalies in two and higher dimensions has been
carried out in Ref. \cite{ludwig}; 
these results have identified some $\Z$ classes of
topological states in $d$ dimensions that actually fit within the
tenth-fold classification of non-interacting topological systems 
\cite{kitaev}, thus extending it to interacting cases.

The next step is to discuss the Laughlin argument for the non-chiral
topological state of the quantum spin Hall effect (see Fig. \ref{fig1}(b)).
Consider the system made of two copies of the $\nu=1$ Hall effect
having opposite spin and chiralities. 
The time reversal (TR) transformation ${\cal T}$ acts on up 
and down spin electrons, resp. $\psi_{\uparrow}$ and 
$\psi_{\downarrow}$, as follows,
\be
{\cal T}:\  
\psi_{k\uparrow} \to \psi_{-k\downarrow} \ ,
\qquad 
\psi_{k\downarrow} \to -\psi_{-k\uparrow},
\label{TR-action}
\ee  
thus leaving the system invariant.

\begin{figure}[t]
\begin{center}
\includegraphics[width=12cm]{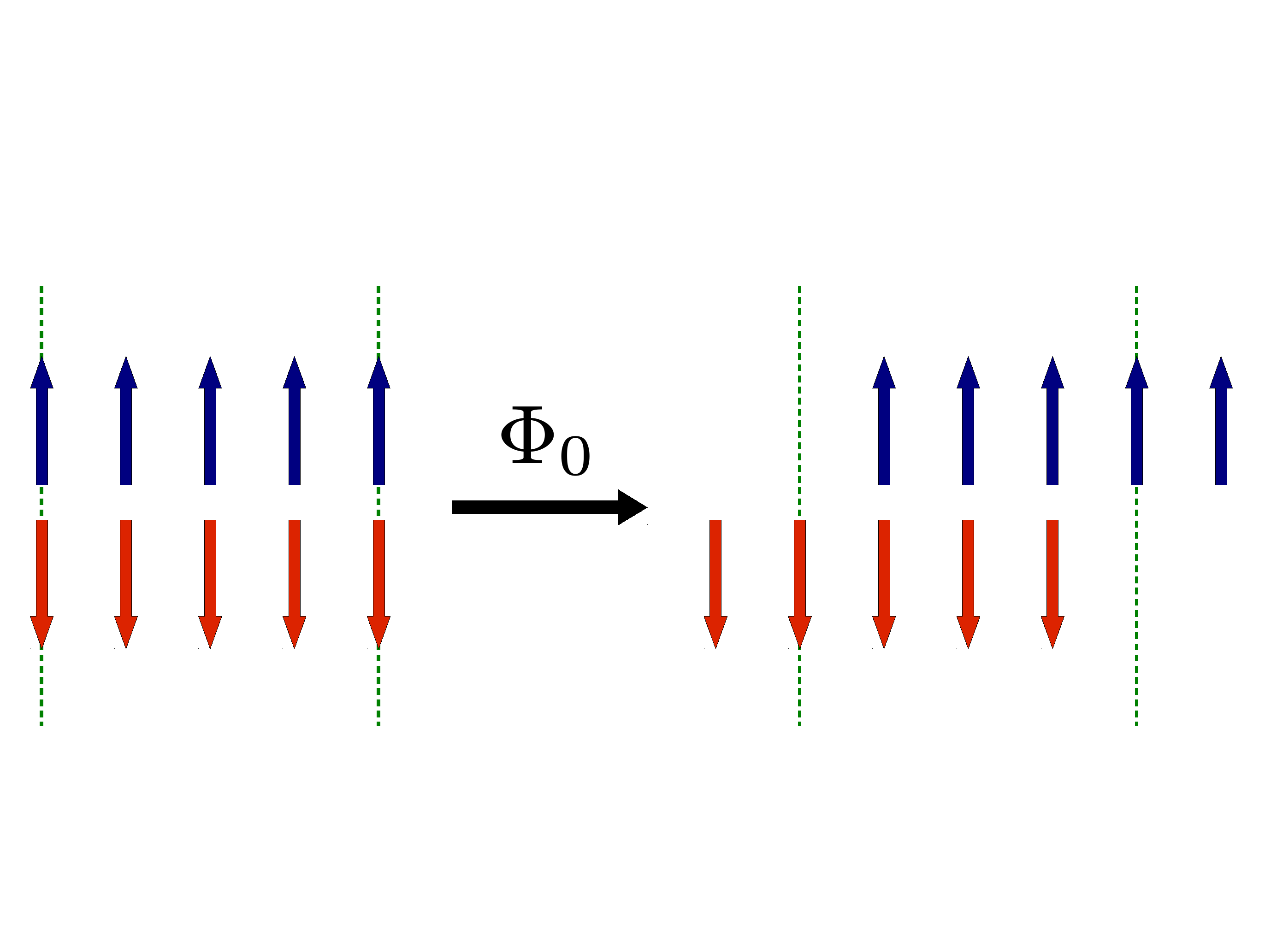}
\caption{Flux insertion in the QSHE: up and down spins are displaced
  w.r.t. the Fermi surfaces at $L$ and $R$ edges of the annulus
(dashed-dotted lines).}
\label{fig2}
\end{center}
\end{figure}

The addition of one flux causes the drift of up and down electrons in opposite
directions with respect to the Fermi surface at each edge (Fig. \ref{fig2}).
From the point of view of the CFT at one edge, say the outer one,
the effect is to create a neutral excitation with spin one \cite{qz}:
\be
\Delta Q = \Delta Q^\uparrow +
\Delta Q^\downarrow =0, \qquad 
\Delta S =\frac{1}{2} -
\left(-\frac{1}{2}  \right)=1,
\label{one-flux-qshe}
\ee  
where $\Delta S=\Delta Q^\uparrow=\nu^\uparrow$, i.e. the spin Hall current 
is equal the Hall current of one chiral component (in appropriate units).

Arguing as in the chiral case, we can characterize this topological
state by the spin chiral anomaly $U(1)_S$ corresponding to an integer
spin Hall conductivity.
However, the quantum spin Hall effect is a rather academic model of
topological insulator: in general, spin-orbit coupling
and other relativistic effects cannot be neglected, that only conserve the
total angular momentum; thus, the $U(1)_S$ symmetry is explicitly broken,
the spin current is not defined and the spin Hall conductivity vanishes.

In the following, we want to discuss general topological insulators
that only possess time reversal symmetry and cannot be characterized
by anomalies of continuum symmetries; these
represent interacting topological phases of a different kind.
In our analysis we shall mostly keep the spin-quantum Hall description
of topological insulators, that can be investigated by CFT methods,
corresponding to the the ideal limit of spin-orbit coupling switched
off.  Nevertheless, we shall discuss properties that do not rely on
the spin being conserved.


\subsection{Fu-Kane-Mele flux argument}

A topological insulator with a single free fermion edge mode per spin
(chirality) is stable because the mass term coupling the two
chiralities is forbidden by TR symmetry, as follows:
\be
{\cal T}:
H_{\rm int.} = m\! \int\! \psi^\dagger_\uparrow 
\psi_\downarrow + h.c. \ \to\ 
- H_{\rm int.} \ .
\label{mass-term}
\ee 
Another TR invariant mass term can be written that couples two fermions
modes per spin, but a single fermion always remains massless
in a system with an odd number of modes \cite{qz-rev}.
Of course, if TR symmetry is broken all edge excitations become gapful
(and the insulator trivial).

The analysis of gapful instabilities due to more general,
non-quadratic interactions compatible with TR can be done in some
cases, but we consider here another criterion for stability that is
associated with a symmetry and a discrete $\Z_2$ anomaly. 
This is the Fu-Kane-Mele flux
insertion argument called the ``spin pump'' (a cyclic adiabatic
process) \cite{kane}.  We mostly follow the presentation of Ref.\cite{ls}.

The insertion of magnetic flux breaks TR symmetry, owing to:
\be
{\cal T}H\left[\Phi\right]{\cal T}^{-1} =
H\left[-\Phi\right] \ .
\label{TR-ham}
\ee  
This relation together with the periodicity $H[\Phi]=H[\Phi+\Phi_0]$, implies 
that the bulk Hamiltonian is TR invariant for a discrete set of flux values:
\be
\Phi=0,\frac{\Phi_0}{2},\Phi_0,
\frac{3\Phi_0}{2},\dots .
\label{half-flux}
\ee  

The Fu-Kane-Mele analysis  of band insulators let them
to define an index called ``TR invariant polarization'',
$(-1)^{P_\theta}=\pm 1$, that enjoys the following properties:

i) It is a bulk topological quantity, conserved by TR symmetry.

ii) Its value is equal to the spin parity (fermion number) at the edge,
\be
(-1)^{P_\theta}= (-1)^{N_\uparrow+N_\downarrow}=(-1)^{2S}.
\label{s-p}
\ee  

iii) In a stable topological insulator, it changes value between TR
invariant points (\ref{half-flux}) separated by half flux
$\Delta\Phi=\Phi_0/2$.

\begin{figure}[t]
\begin{center}
\includegraphics[width=14cm]{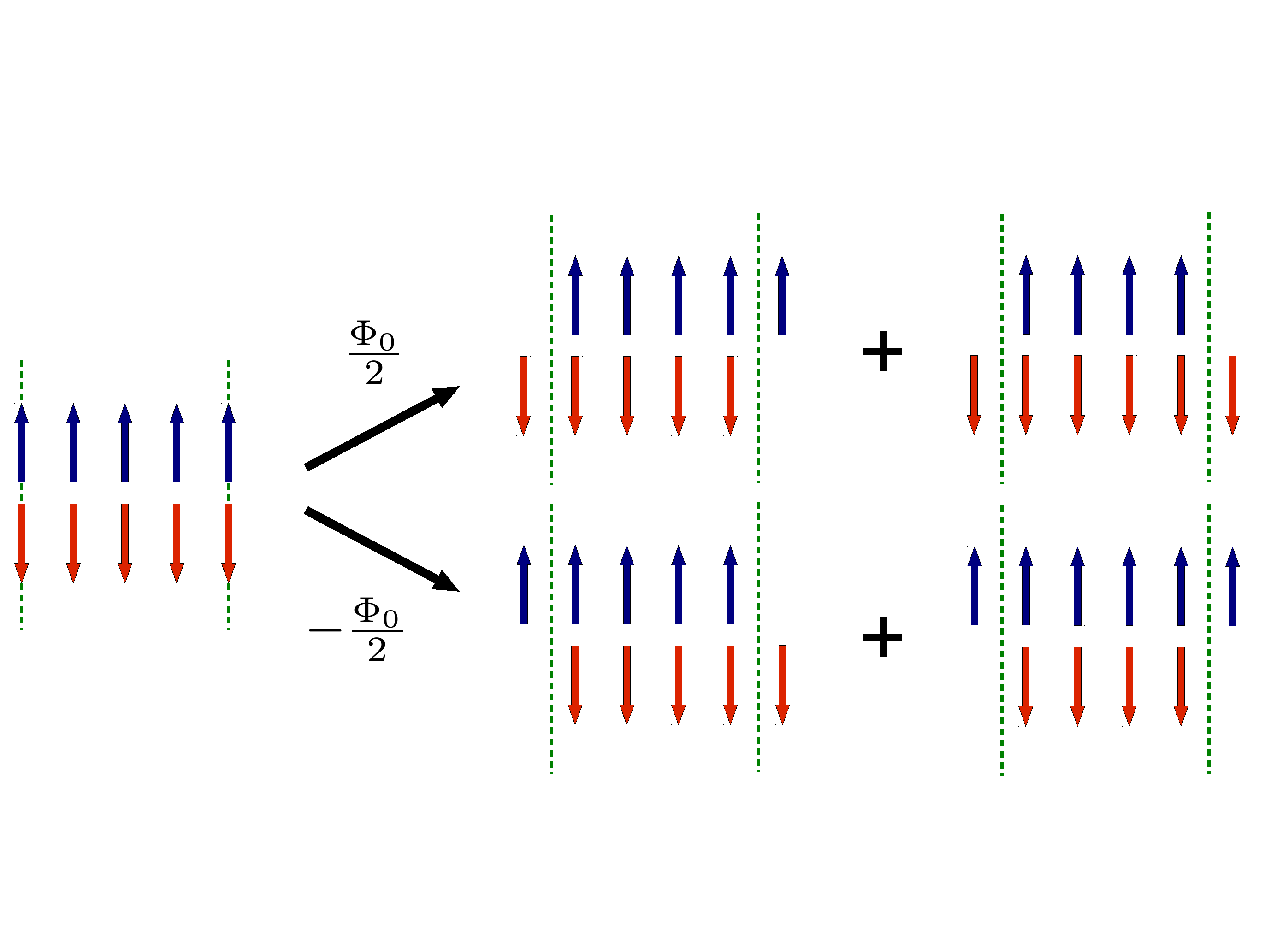}
\caption{Flux insertions in the QSHE creating a spin
one-half excitation at each edge. Their local TR partners are also included.}
\label{fig3}
\end{center}
\end{figure}

The stability argument goes as follows: adding $\pm\Phi_0/2$
fluxes in the center of the annulus creates a spin
one-half excitation at both boundaries, the two cases being related by a
TR transformation (see Fig. \ref{fig3}). Each of these excitations
 has a TR partner locally at the boundary, making a total of four
states.  Next, one invokes the Kramers theorem at the TR-invariant
point $\Phi=\Phi_0/2$ to argue that the two spin one-half excitations
at one edge are degenerate in energy and orthogonal, this degeneracy
being robust to any TR symmetric perturbation\footnote{
The local version of the Kramers theorem at each edge of the system
has been discussed in \cite{ls}.}.

\begin{figure}[t]
\begin{center}
\includegraphics[width=10cm]{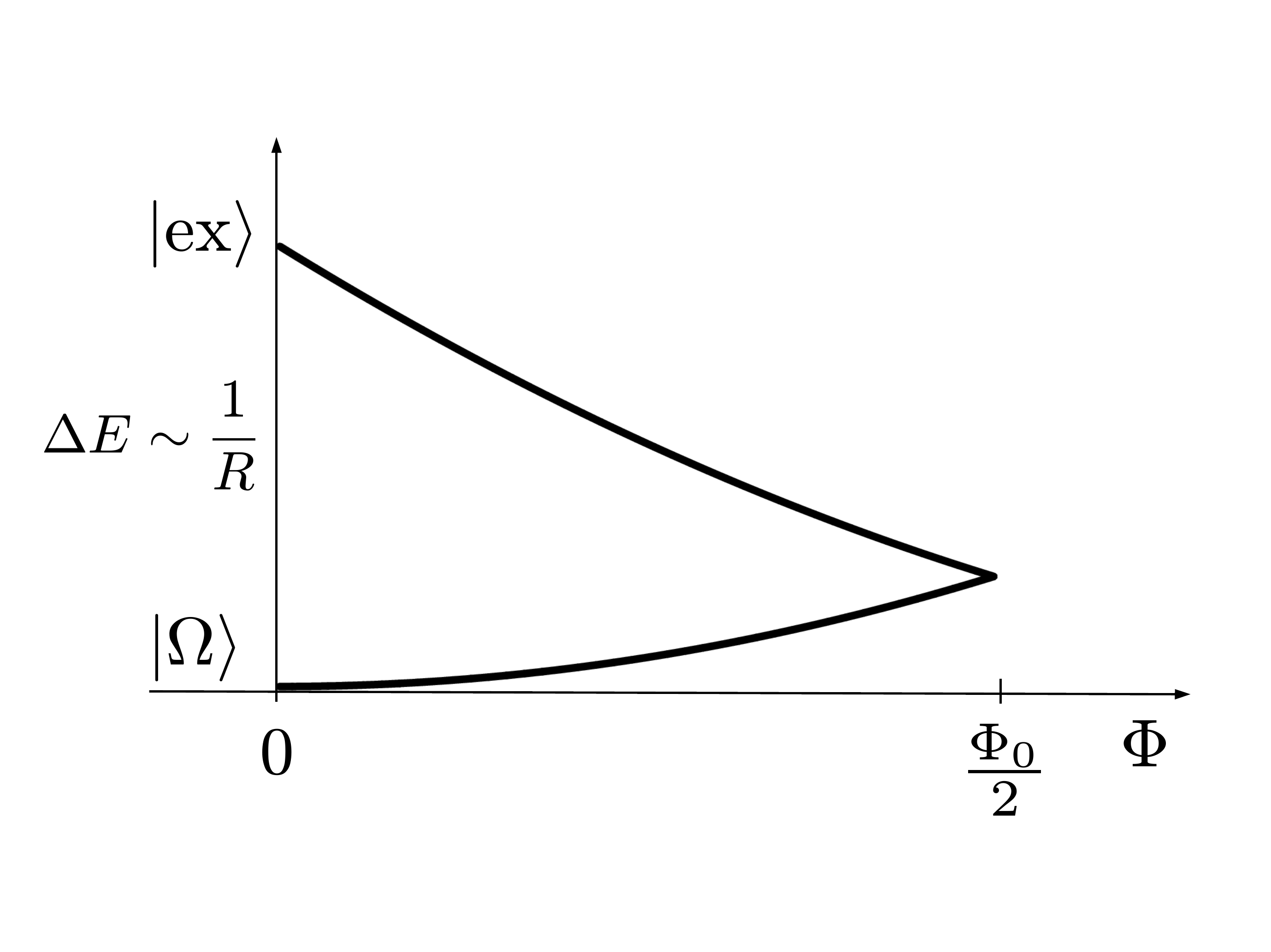}
\caption{Kramers degeneracy at half flux quantum.}
\label{fig4}
\end{center}
\end{figure}

The energy change of the edge ground state as the flux is varied from zero
to $\Phi_0/2$ is shown in Fig. \ref{fig4}. At $\Phi_0/2$, the evolved ground
state $\vert\Omega\rangle$ necessarily meets with one excited state
$\vert {\rm ex}\rangle$ owing to Kramers theorem.
Going back to $\Phi=0$, the excited state must
have an energy $O(1/R)$ from the work done by adding a flux quantum
in a system of size $R$. 
It then follows that the existence of a Kramers (spin one-half) pair at the
edge for $\Phi=\Phi_0/2$ implies the presence of a gapless excitation
at $\Phi=0$ that is protected by TR symmetry.
In the case of two fermion modes, the corresponding spin one excitation
created at the boundary would not be protected by Kramers theorem
against energy splitting from its TR companions.
The argument then extend to odd and even numbers of fermion modes.

This completes the argument for stability of the topological phase with
an odd number of fermion modes, leading to the $\Z_2$ classification
of topological insulators in the free fermion case.
Let us add some remarks:

i) The existence of a Kramers pair is signalled by the change
of spin parity of the ground state upon adding half flux,
\be
\Phi=0: \ (-1)^{2S} =1 \ \  \longrightarrow \ \
\Phi=\frac{\Phi_0}{2}:\ (-1)^{2S}=-1.
\label{change s-p}
\ee

ii) The spin parity is conserved by TR symmetry, being just another
way to state the Kramers theorem. This $\Z_2$ invariance is the
remnant of the continuous $U(1)_S$ symmetry of the quantum spin Hall
effect that gets broken by relativistic effects.

iii) At the two TR symmetric points, $\Phi=0,\Phi_0/2$, the spin
parity takes different values without having included TR breaking
terms in the Hamiltonian. Therefore, this quantity is anomalous.

In conclusions, TR invariant topological insulators are associated with the
$\Z_2$ spin parity symmetry $(-1)^{2S}$ that is anomalous. 
The full spin symmetry $U(1)_S$ would also be anomalous but 
it is explicitly broken in general topological insulators.
We note that the Fu-Kane-Mele argument is 
a generalization of the Laughlin argument that makes it apparent
the $\Z_2$ spin parity anomaly\footnote{
Discrete anomalies in topological insulators have also been discussed
in \cite{stern}; for a general introduction, see \cite{anom}.}.
The goal of this paper is to generalize the stability analysis
to interacting systems through the study of partition functions.

\begin{figure}[t]
\begin{center}
\includegraphics[width=10cm]{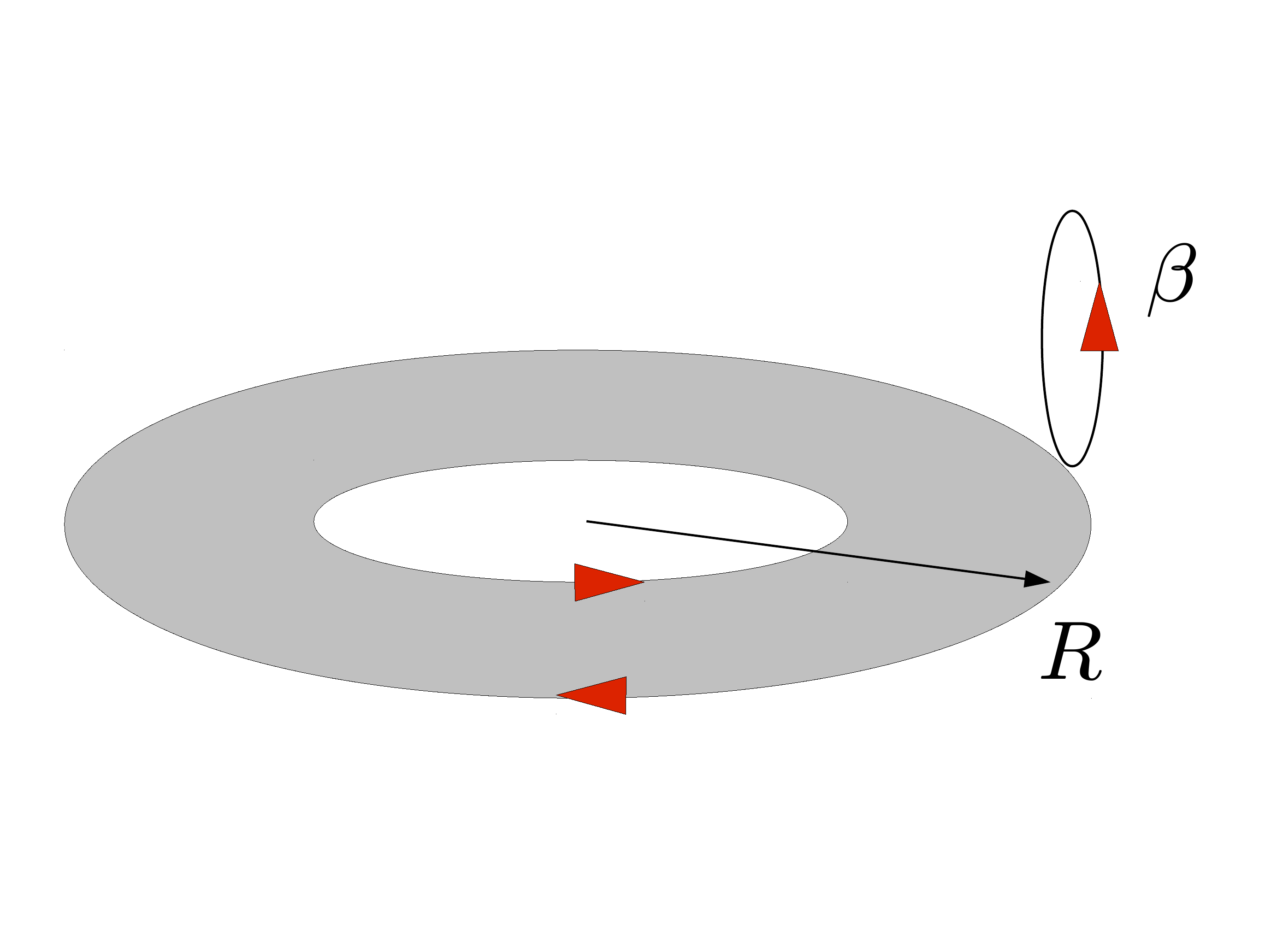}
\caption{Torus geometry with periods $(2\pi R,\b)$.}
\label{fig5}
\end{center}
\end{figure}


\section{Partition functions of topological insulators}

\subsection{Chiral edge system}

We first recall from Ref. \cite{cz-mod} the construction 
of the partition function
of the quantum Hall effect for the Laughlin states, $\nu=1/p$, $p$ odd;
next, we generalize it to topological insulators and discuss
the $\Z_2$ anomaly and the stability criterion in this setting.

We consider the grand-canonical partition function of states
at the outer edge of the annulus with radius $R$ (see Fig. \ref{fig5}). 
This circle and the time period $\b$ realize the geometry of the torus.
The energy and momentum of chiral excitations are 
expressed by the eigenvalues of the Virasoro generator $L_0$,
$E=P=L_0/R$ (we set the Fermi velocity $v=1$ for simplicity).
The trace over the states with Gibbs weight involving chemical 
and electric potential decomposes into orthogonal sectors 
${\cal H}^{(\l)}$, corresponding
to a given value of fractional charge plus any number of electrons,
$Q=\l/p+n$, $n\in\Z$; there are $p$ sectors for $\l=0,1,\dots,p-1$.

The partition function for one sector takes the form:
\ba
K_\l(\t,\z;p)&=&{\rm Tr}_{{\cal H}^{(\l)}}\!
\left[ \exp\left(i2\pi \t L_0 + i2\pi \z Q\right) \right]\nl
&=&\frac{F(\t,\z)}{\eta(\tau)}
\sum_{n\in\Z} \exp\left(i2\pi\left( \t \frac{(np+\l)^2}{2p} + 
\z \frac{np+\l}{p}\right) \right);
\label{pf}
\ea
this function is parameterized by the two complex numbers,
\be
\t= \frac{i\b}{ 2\pi R}+ t, 
\qquad
\z =\frac{ \b}{ 2\pi} (iV_o +\mu),
\label{tau-zeta}
\ee
that are the modular parameter $\t$ and the
``coordinate'' $\z$. $\I \t>0$ is the ratio of the two periods and 
$\R \t$ is the torsion parameter conjugate to momentum $P$;
$\z$ contains the electric $V_o$ and chemical $\mu$ potentials.

The expression of $K_\l$ (\ref{pf}) involves a sum of characters of the
representations of the $U(1)$ current algebra of the $c=1$ CFT with
charges $Q=\l/p+n$ and conformal weight $h=(\l+pn)^2/2p$, as is
apparent in the exponent of (\ref{pf}) \cite{cft}.  
It can be obtained by canonical
quantization \cite{cdtz} as well as by using CFT representation 
theory supplemented
by some physical conditions on the charge and statistics of electron
excitations \cite{cz-mod}.  
The formula involves the ratio of a theta function with
characteristics 
$\Theta\big[\substack{\lambda/p\\0} \big] (\zeta|p\tau)$ 
and the Dedekind function $\eta(\t)$ describing
particle-hole excitations.  The non-holomorphic prefactor
$F=\exp\left[-\pi (\I z)^2/p\,\I \t \right]$ is explained in
Ref.\cite{cz-mod}. Altogether, the partition function of one chiral edge is
given by the multiplet of functions $K_\l$, for $\l=1,\dots,p$,
having periodicity $K_{\l+p}=K_\l$, that correspond to the $p$ anyon sectors.

The torus geometry is left invariant by the modular transformations, the
discrete coordinate changes consistent with the double periodicity.
These act on the modular parameter $\t$ and the coordinate $\z$ as follows
\cite{cz-mod}:
\be 
\t\ \to \ \frac{a\t +b}{c\t +d}, \qquad \z \ \to \ \frac{\z}{c\t +d}, \qquad
a,b,c,d \in \Z,\qquad ad-bc=1,
\label{mod.tr}
\ee
and span the group $\G =SL(2,\Z)/\Z_2$. The modular group is 
infinite dimensional and is generated by two transformations,
$T : \t \to \t+1,\ \z \to\z$ and 
$S:\t \to -1/\t,\ \z\to -\z/\t$, obeying the relations
$S^2=(ST)^3=C$, where $C$ is the charge conjugation matrix,
$C^2=1$ \cite{cft}.
In addition, there are the two periodicities of the coordinate $\z$
at $\t$ fixed: $U:\z \to \z+1$ and $V:\z \to \z+\t$.

The modular transformations belong to the group
of two-dimensional diffeomorphisms of the torus, being the global
transformations not connected to the identity:
they are the ``large'' gauge transformations of the conformal theory
placed in a gravitational background. 
In similar way, the flux insertions are large gauge transformations
of the electromagnetic background.
The lack of invariance of the partition function signals the presence of 
gravitational and gauge anomalies, respectively.
These do not led to inconsistencies if the backgrounds are 
classical, i.e. not quantized, as in our case, but nevertheless
characterize the low-energy physics of the system \cite{kitaev}. 

The multiplet of $K_\l$ transforms linearly under the modular group
and each generator has physical significance, as we now recall \cite{cz-mod}.  
The $S$ transformation reads:
\be
S:\ \  K_\l\left(\frac{-1}{\t},\frac{-\z}{\t}\right) = e^{i\varphi}\
\sum_{\mu=1}^p S_{\l\mu} K_\mu(\t,\z),\qquad
S_{\l\mu}=\frac{1}{\sqrt{p}}\exp\left(i2\pi \frac{\l\mu}{p}\right),
\label{S_transf}
\ee where $S_{\l\mu}$ is the modular $S$-matrix and $\varphi$ is an
overall phase. As is well know in the CFT literature, the $S$ unitary
transformation expresses a completeness condition on the spectrum of the
theory; it also determines the fusion rules of excitations via the 
Verlinde formula \cite{cft}.

The invariance under $T^2$ transformation up to a global phase,
\be
T^2 :\ \ K_\l(\t+2,\z) =
\exp\left(i4\pi h_\l\right) K_\l (\t,\z), \qquad h_\l=\frac{\l^2}{2p},
\label{T2_transf}
\ee
implies that the electron excitations being summed up
in each anyon sector have odd integer statistics (half integer conformal
dimension). On the other hand, the $U$ invariance,
\be
U :\ \ K_\l(\t,\z+1)= \exp\left(i2\pi \l/p\right)  K_\l (\t,\z),
\label{U_transf}
\ee 
implies electrons with integer charge. The $T^2$ and $U$ transformations
are readily checked from the expression (\ref{pf}).

Finally, the $V$ transformation,
\be
V :\ \ K_\l(\t,\z+\t) = e^{i\f} K_{\l+1} (\t,\z),
\qquad\quad \D\Phi=\Phi_0,
\label{V_transf}
\ee
(with $\f$ another global phase) is very important because
it realizes the change of electric potential following the addition of one
flux quantum $\Phi_0$. The shift of index $\l\to\l+1$ 
corresponds to the spectral
flow $Q\to Q+\nu$, $\nu=1/p$, discussed in the previous section.

Altogether the single edge is described by a multiplet of partition
functions $K_\l$, that is not modular invariant, meaning that the
chiral anomaly, i.e. the spectral flow, implies a discrete
gravitational anomaly.  More precisely, the $K_\l$ transform in a
linear unitary vector representation of $\G_\th={\rm span}(T^2,S)$, a
subgroup of $\G$ including only $T^2$ transformations and thus allowing
fermionic statistics \cite{cz-mod}.  The dimension $p$ of the
representation is equal to the value of the Wen topological order
\cite{wen-book}.


\subsection{Non-chiral edge}
The partition function for the quantum spin Hall system made by
a pair of Laughlin states is obtained by
combining chiral and antichiral sectors for up and down spins,
respectively, thus obtaining the expressions
$K_\l^\uparrow\; {\ov K}_\mu^\downarrow$.
In the case of the Hall effect of charge and spin, the presence of 
extended states in the bulk of the annulus allows for the matching 
of fractional charges between opposite edges. 
In the case of topological insulators, the bulk is insulating,
thus the fractional charges should be matched locally at each edge
(the $U$ condition). 
We thus obtain the following expression for the partition function 
of a single edge:
\be
Z^{NS}\left(\t,\z \right)=
\sum_{\lambda=1}^p K^\uparrow_\l\; {\ov K}^\downarrow_{-\l}\ .
\label{NS-pf}
\ee
 This is invariant under $S,T^2,U,V$.
It turns out that this quantity is formally equal to the quantum Hall
effect partition function for the system of two edges, i.e. for the
whole annulus \cite{cz-mod}.
However, the physical interpretation in the case of topological insulators
is rather different, since it only describes a single edge of the annulus
and the different charge sectors do not correspond to bulk anyonic
excitations, but just describe degenerate ground states 
at one edge.

The expression $Z^{NS}$ is not completely modular invariant, because
the transformations in the quotient $\G/\G^\th\sim S_3$ deform it into
other quantities according to a general pattern.  Indeed,
the partition functions of fermionic systems always involve four
terms corresponding to the four spins structures needed for
defining spinors on the torus \cite{gso}. These amount to choosing periodic
$(P)$ and antiperiodic $(A)$ boundary conditions for fermion fields in each
direction (in general there are $2^{2g}$ terms on a genus $g$
surface). These terms are known as the Neveu-Schwarz $(NS)$ and Ramond $(R)$
sectors and their tildes, as follows:
\be 
NS, \ \wt{NS},\ R,\ \wt{R}, \quad {\rm
  respectively:}\quad (AA), \ \ (AP),\ \ (PA),\ \ (PP).
\label{spin.struct.}
\ee
The expression (\ref{NS-pf}) is identified with the Neveu-Schwarz sector
since the natural fermionic boundary conditions are antiperiodic:
\be
Z^{NS}=\Tr_A\left[\exp\left(i2\pi\t L_0+i2\pi \z Q + h.c.\right)\right].
\label{pf-trace}
\ee
The other expressions are defined as:
\ba
Z^{R}&=&\Tr_P\left[\exp\left(i2\pi\t L_0+i2\pi \z Q + h.c.\right)\right],
\nl
Z^{\wt{NS}}&=&\Tr_A\left[(-1)^{N_\uparrow+N_\downarrow}
\exp\left(i2\pi\t L_0+i2\pi \z Q + h.c.\right)\right],
\nl
Z^{\wt{R}}&=&\Tr_P\left[(-1)^{N_\uparrow+N_\downarrow}
\exp\left(i2\pi\t L_0+i2\pi \z Q + h.c.\right)\right],
\label{pf2-trace}
\ea
where periodic  conditions in time introduce the sign 
$(-1)^F=(-1)^{N_\uparrow+N_\downarrow}$.

The modular transformations among the four terms are depicted
in Fig. \ref{fig6}(b) and are explicitly checked in the Appendix.
They form a triplet, $Z^{NS},Z^{\wt{NS}},Z^R$, and a singlet, $Z^{\wt{R}}$.
Each one of the four spin sectors is made of $p$ ``anyonic'' sectors: 
\be
Z^s=\sum_{\l=1}^p K^{\uparrow s}_\l\ov{K}^{\downarrow s}_{-\l},
\qquad s=NS, \wt{NS}, R,\wt{R}.
\label{pfc1}
\ee
Furthermore, these $Z^\s$ are invariant under $T^2$, i.e. possesses
fermions excitations, under $U$ for charge matching and 
under $V$ for the spectral flow among the $p$ anyon sectors.

Regarding the $\wt{NS}$ sector, the partition function and 
its anyon sectors are defined as follows,
\ba
Z^{\wt{NS}}\left(\t,\z\right)&=& Z^{NS}\left( \t+1,\z\right),
\nl
K^{\wt{NS}}_\l\left(\t,\z\right)&=& 
e^{i\theta_\l}\ K_\l\left(\t+1,\z\right).
\label{def_NSt}
\ea
In this expression, the phase 
$\theta_\l=2\pi\left(\frac{\l}{2}-\frac{\l^2}{2p} +\frac{1}{24}\right)$
is included for convenience; we also write $K_\l^{NS}\equiv K_\l$
and omit spin arrows for simplicity.
Note that the term brought by the $T$ transformation 
in the summation over fermions inside $K^{\wt{NS}}_\l$,
i.e. $e^{i2\pi L_0}$, is proportional
to  $(-1)^F$ owing to the half integer conformal dimension of fermions.

The Ramond sector is similarly obtained by acting with $ST$ on $Z^{NS}$ and
$Z^{\wt{R}}$ is defined by inserting the $(-1)^F$ sign into the Ramond
expression. Explicit examples will be given later and are collected in the
Appendix. Altogether, the pattern in Fig. \ref{fig6}(b) shows the response of
the topological insulator edge to modular transformations, i.e discrete
diffeomorphisms not connected to the identity.  The four spin
sectors, each one made of $p$ anyon sectors, realize a kind of decoupling of
anyonic and fermionic properties, as it will be more clear in the following.
 
\begin{figure}[t]
\begin{center}
\includegraphics[width=15cm]{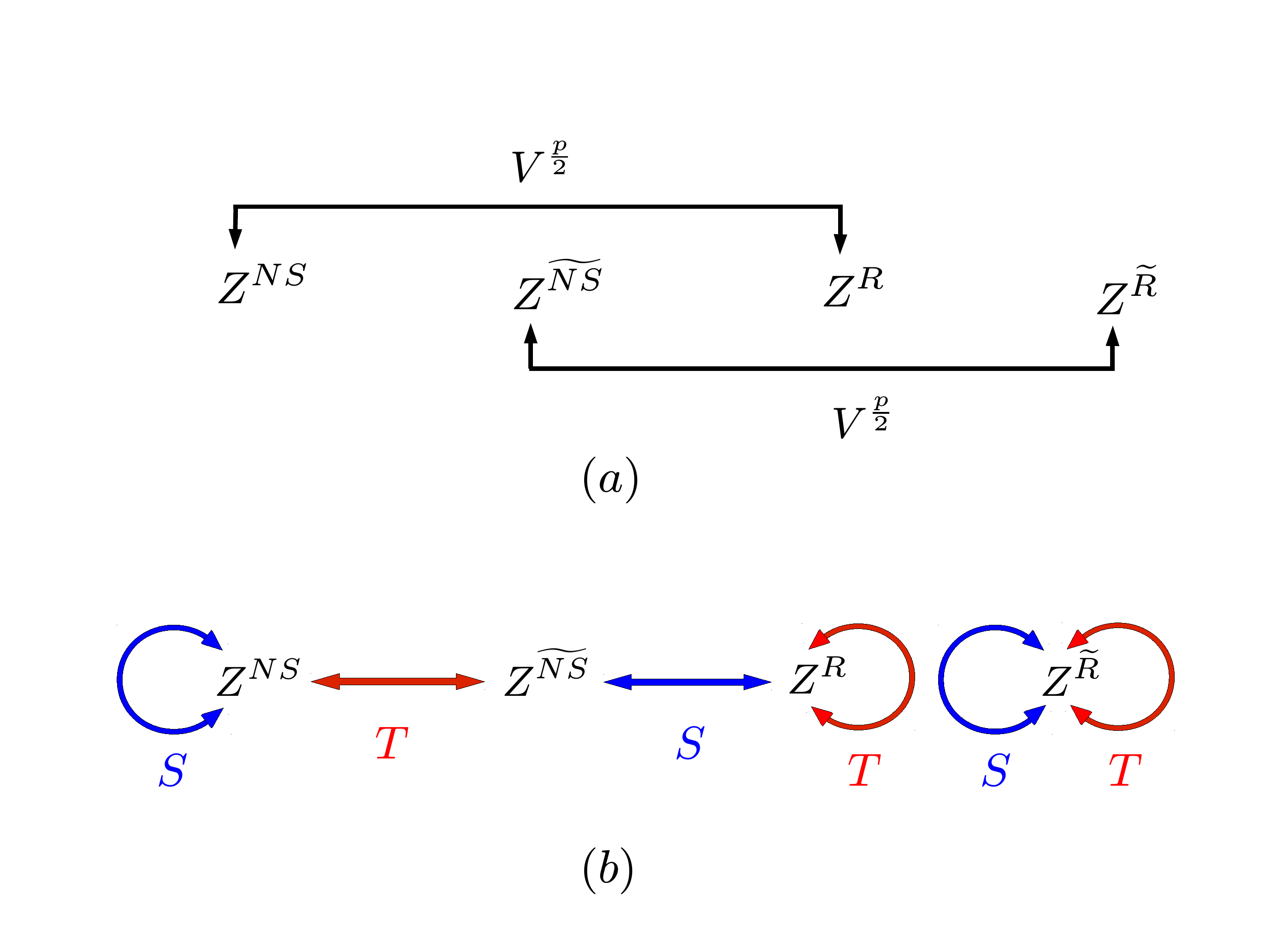}
\caption{Actions of (a) $p/2$ flux insertions and (b) 
modular transformations on the four spin sectors 
$Z^{NS}, Z^{\wt{NS}}, Z^R, Z^{\wt{R}}$ (odd $p$ case).}
\label{fig6}
\end{center}
\end{figure}

\subsection{Stability analysis}
We now consider the response to adding magnetic fluxes and recover the
Fu-Kane-Mele stability analysis in the context of partition functions.  
As already said, each spin
sector is invariant under $V$, the addition of one flux quantum.  The addition
of half flux $V^{1/2}$ transforms the sectors as shown in Fig. \ref{fig6}(a).  
In particular, the Neveu-Schwarz sector is mapped into the Ramond sector,
\be
V^{\frac{1}{2}}: Z^{NS}\left(\t,\z\right)\ \to \ 
Z^{NS}\left(\t,\z +\frac{\t}{2} \right)=
Z^R\left(\z,\t\right).
\label{NStoR}
\ee 
The addition of any half integer number of fluxes $V^{\frac{1}{2}+n}$
yields the same result, up to a reshuffling of anyon sectors within $Z^R$
(cf. Eq.(\ref{pfc1})).  
In order to disentangle the anyonic degeneracy from the electron degeneracy
relevant for the Fu-Kane-Mele stability analysis \cite{kane}, 
we follow th extention of the argument due to Levin-Stern \cite{ls}.
We observe that the addition of $p$ fluxes creates an electron excitation
within the same anyon sector, as it corresponds to a symmetry of each
$K_\l$,
\be
V^p: K_\l\ \to\ 
K_{\l+p}=K_\l, \qquad\quad \Delta Q^\uparrow=\frac{p}{p}=1 \ .
\label{p.fluxes}
\ee

Therefore, the addition of $p/2$ fluxes will create a spin one-half
excitation in the topological insulator edge,
 $\Delta S =\Delta Q^\uparrow=1/2$,
while staying in the same anyonic sector.
It is therefore convenient to define the Ramond sectors by
the action of $V^{\frac{p}{2}}$:
\be
V^{\frac{p}{2}}: \ 
K_\l\left(\t,\z\right)\ \to K_\l\left(\t,\z+\frac{p\t}{2}\right)\sim
K_{\l+\frac{p}{2}}\left(\t,\z\right)= K^R_\l\left(\t,\z\right),
\label{def_R}
\ee
where $(\sim)$ stands for equality up to a global phase.

We can now extend the stability analysis discussed in the 
free fermion case (Section two).
Upon applying $p/2$ fluxes, the Neveu-Schwarz ground state
$\left\vert\Omega\right\rangle_{NS}$, 
the lowest state in $K_0\ov{K}_0$, evolves in the Ramond ground state 
$\left\vert\Omega\right\rangle_R$ present
in $K^R_{0}\ov{K}_0^R=K_{p/2}\ov{K}_{p/2}$. 
This becomes a spin-half partner of a degenerate Kramers pair: following
the argument of Section two, this proves
the stability of the topological phase in the system made by 
a pair of Laughlin states, for any $p$ value. 

The existence of the Kramers pair and the behaviour of the spectrum
shown in Fig. \ref{fig4} can be easily checked by inspecting the low
lying states contained in $K_{0}\ov{K}_0$ and $K^R_{0}\ov{K}_0^R$: one
expands to lowest order in $q\ov{q}$, with $q=\exp(i2\pi\t)$, and
checks the terms to $O(w^0\ov{w}^0)$, $w=\exp(i2\pi\z)$, i.e. not
involving additional particles.

We can also recompute the spin parity of the Neveu-Schwarz
and Ramond ground states:
\be
(-1)^{2S}\left\vert\Omega\right\rangle_{NS}=
\left\vert\Omega\right\rangle_{NS}
\ \to\ 
(-1)^{2S}\left\vert\Omega\right\rangle_R
=- \left\vert\Omega\right\rangle_R,
\qquad
\Delta S =\Delta Q^\uparrow= \frac{1}{2}\ .
\label{sp.groundst}
\ee
The validity of the Levin-Stern stability index discussed in the 
Introduction (Eq.(\ref{ls-index})) is thus verified for this system:
\be
2\Delta S=2\Delta Q^\uparrow=
\frac{\sigma_{sH}}{e^*}=1\ ,  \qquad (-1)^{2\Delta S}=-1,
\label{ls-index2}
\ee 
where $\sigma_{sH}=\nu^\uparrow=1/p$ and $1/e^*=p$ is the number of
charge sectors, i.e. the periodicity of $K_\l$.
The (would-be) spin transport $\D S$ involved in this index, equal to
the Hall current of one chiral component, is that relative to the
addition of half of the fluxes needed for creating an electron excitation 
within a given anyon sector.

In conclusion, we have found the important fact that the spin parity of the
Ramond ground state is different from that of the Neveu-Schwarz ground state.
This is a manifestation of the discrete anomaly $\Z_2$: different sectors of
the path integral (Eq.\eqref{pf-trace} and \eqref{pf2-trace}) 
have associated different quantum numbers \cite{anom}; in the same way,
the chiral sectors $K_\l$ have associated different charges due to the
$U(1)_Q$ anomaly (Eq.(\ref{V_transf})).


\subsection{Stability and modular non-invariance}

In a fermionic non-chiral system composed of the four spin
sectors (\ref{pf-trace}),(\ref{pf2-trace}) it is always possible 
to find a modular invariant
partition function by summing over all sectors,
\be
Z_{\rm Ising}= Z^{NS}+Z^{\wt{NS}}+ Z^R+Z^{\wt{R}}.
\label{Ising}
\ee
This is the so-called Ising projection because it occurs in CFTs
applied to statistical models like the Ising model, its supersymmetric
generalizations etc. \cite{cft}. 
This quantity is $S,T,U,V^{\frac{1}{2}}$ invariant.

Nonetheless, we would like to argue that the theory defined by $
Z_{\rm Ising}$ may not be consistent with the TR symmetry of
topological insulators, implying spin parity conservation.  In
presence of the $\Z_2$ anomaly, the partition function (\ref{Ising})
sums spin sectors with different values of the ground state spin
parity, thus violating TR symmetry.  Therefore, this symmetry is
explicitly broken or not defined in that theory.

If we insist on preserving TR symmetry, we should not
sum over spin sectors and leave them as the components of a 
four-dimensional vector,
\be
Z_{\rm TR}=\left( Z^{NS}, Z^{\wt{NS}}, Z^R, Z^{\wt{R}}\right),
\label{ZTR}
\ee 
that carries a non-trivial representation of $\G/\G_\th\sim S_3$.
Therefore, the $\Z_2$ anomaly is associated to a $S_3$ gravitational
anomaly.  In the theory described by this set of partition functions,
$Z^{NS}$ represent the TR invariant edge system, while the other
functions, $Z^{\wt{NS}}, Z^R, Z^{\wt{R}}$, are excited states of the
system in presence of electromagnetic or gravitational backgrounds.

We thus obtain the following result, later shown to be valid
in general:
\ba
&&\!\!\!\!\!\!\!\!\!\!\!\!\!\!\!\!\!
 TR \ symmetry + anomaly \  \leftrightarrow\ 
 no\ modular\ invariance \  \leftrightarrow\ 
 topological\  insulator,
\nl
&&\!\!\!\!\!\!\!\!\!\!\!\!\!\!\!\!\!
 TR \ symmetry +modular\ invariance \ \leftrightarrow\ 
 no\ anomaly \  \leftrightarrow\ 
trivial\  insulator.
\label{dog}
\ea

Let us add some remarks:

i) As in the case of the quantum Hall effect, the anomaly can be cancelled
globally on the whole system by combining the partition functions of the two
edges of the annulus, leading to a global modular invariant.

ii) In the quantum Hall state, the chiral partition functions like
$K_\l$ cannot be combined into a modular invariant for a single edge
(unless a special case with $c=24$).  In the topological insulator,
they can or cannot be combined depending on the fate of the $\Z_2$
symmetry; this is a manifestation of the symmetry protection of the
topological phase, because this can became trivial if the symmetry is
not enforced. A related statement is that the stability of these
systems is not solely determined by CFT properties but by the way
$2+1$-dimensional bulk symmetries are attached to CFT edge states.

iii) In the $c=1$ Laughlin spin state considered in this
section, the two partition functions $Z_{\rm TR}$ and 
$Z_{\rm Ising}$ can be obtained by canonical quantization of the 
following Lagrangians. $Z_{NS}$ is obtained by independent quantization
of two copies of the chiral boson theory  \cite{cdtz},
\be
S_1=\frac{1}{4\pi}\int \partial_x\x\left(\partial_t\x -v \partial_x\x \right)
+\frac{1}{4\pi}\int \partial_x\c\left(\partial_t\c +v \partial_x\c \right),
\label{}
\ee
with rational compactified radius $r^2=p/q$, $p,q \in\Z$, and  antiperiodic
boundary conditions.  The spectrum of Virasoro states is given 
by $h=n^2/2pq$ in one chiral theory and  $\ov{h}=m^2/2pq$ in the other one,
with $n,m\in\Z$.

The partition function $Z_{\rm Ising}$ is obtained by the quantization
of the non-chiral compactified boson \cite{cft},
\be
S_2=\frac{1}{4\pi}\int\left(\partial_t\phi\right)^2 -v^2 
\left(\partial_x\phi \right)^2.
\label{}
\ee
One obtains the standard spectrum 
$(h,\ov{h})=\left(\left(\frac{n}{2r}+mr\right)^2/2, 
\left(\frac{n}{2r}-mr\right)^2/2 \right)$, with $r^2=p/2q$,
that spans a two-dimensional even self-dual Lorentzian lattice,
characteristic of compactified bosonic theories \cite{modinv} \cite{gannon}.
The partition function resulting by summing over the lattice is 
modular invariant and can be rewritten in the form (\ref{Ising}):
\be
Z_{\rm Ising}=\frac{1}{\vert\eta(\t)\vert^2}\ \sum_{n,m\in \Z^2}
q^{\frac{1}{2}\left(\frac{n}{2r}+mr\right)^2}\ 
\ov{q}^{\frac{1}{2}\left(\frac{n}{2r}+mr\right)^2}.
\label{selfdual}
\ee
Of course, the choice of other boundary conditions in the theory $S_1$ 
leads to the other three sectors,  $Z^{\wt{NS}}, Z^R, Z^{\wt{R}}$.
The Ising projection then corresponds to imposing the condition
$(-1)^{N^\uparrow +N^\downarrow}=1$ that eliminates fermionic
excitations from both the $(P)$ and $(A)$ spectrum.  
Namely, $Z_{\rm Ising}$ does not describe fermionic excitations and for
this reason it contains less states than $Z^{NS}$.  On the other hand,
the $Z_{\rm Ising}$ possesses additional states of the Ramond  spectrum.
In conclusion, the stable topological system described by $Z^{NS}$
contains single fermion excitations and can be derived from $S_1$;
the unstable, non TR symmetric phase 
has only bosonic states and is naturally described by $S_2$.

iv) Another indication that TR symmetry is not present in $Z_{\rm   Ising}$ 
is given by the violation of the spin-statistics relation in
the Ramond sector. Since $Z^R$ and $Z^{\wt{R}}$ are $T$ invariant, 
they only contain states with integer conformal dimension,
i.e. bosonic statistics. On the other hand, their spin is half
integer, upon summing $S=1/2$ for Ramond ground state to those of
excitations present in $Z^R+Z^{\wt{R}}$, i.e.
$\Delta S=(N^\uparrow +N^\downarrow)/2 \in\Z$.
Therefore, the spectrum does not respect spin-statistics.
On the other hand, the partition function 
$Z'_{\rm Ising}= Z^{NS}+Z^{\wt{NS}}+Z^R-Z^{\wt{R}}$
 would obey the charge-statistics relation and also be modular invariant, 
but it is not invariant under $V^{\frac{1}{2}}$.

\section{General stability analysis and examples}

\subsection{General partition functions}

The conformal theories of general quantum Hall edge states possess not only
charged excitations but also neutral modes that can be Abelian or
non-Abelian. These theories have the affine symmetry $U(1) \times G/H$, where
$U(1)$ is the charge symmetry and $G$ is another (non-Abelian) symmetry
characterizing the neutral part (possibly a coset $G/H$).
The electron field in this theory is represented by the product
of a chiral vertex operator for the charge part and a 
chiral neutral field $\psi_e$ of $G/H$:
\be
\Psi_e=e^{i\a \varphi}\, \psi_e\ .
\label{el-field}
\ee
The field $\psi_e$ should also have Abelian fusion rules with all fields in the
theory: this property is needed for the electrons to have integer statistics
with all excitations and for the ground state wavefunction to be unique, being
written as a CFT correlator\footnote{ 
There are exceptions to the unique identification between wavefunction and
correlators, but they will not be considered here \cite{hansson}.
}.

The field $\psi_e$, called a simple current in the CFT literature
\cite{cft}, can be used to build a modular invariant that couples
neutral and charged parts non-trivially and fulfills the physical
conditions on charge and statistics of the edge spectrum.
The general expression of the partition function for the Hall edge
states obtained in this way is determined uniquely by two inputs: the
choice of neutral $G/H$ theory and of the Abelian field $\psi_e$ that
represents the electron neutral part.  These simple-current
modular invariant partition functions were shown to reproduce earlier
results obtained by physical insight in many models and to build new
ones \cite{cv}.

The construction starts from the one-edge partition sum
for an anyon sector, generalizing the $K_\l$ of the $c=1$ theory
introduced in Section 3.1 (Eq.(\ref{pf})): this involves again a basic anyon
plus any number of electrons added to it, with charge $Q=\l/p+n$,
$n\in \Z$.  It is characterized by $\l$, and the neutral quantum
numbers $(m,\a)$.  In terms of states and fields, the electron
excitations are obtained by fusing the basic anyon field with many
electron fields, always getting a unique output owing to the Abelian
fusion. This conserves the charge $\l$ and another neutral additive
number $m$ of the simple current (assume $m$ modulo $k$ for
simplicity, i.e. a $\Z_k$ neutral charge).  Such partition function
takes the form \cite{cv}: 
\be 
\Th_\l^\a(\t,\z)= \sum_{a=1}^k K_{\l+a p}\left(\t,
k\z; kp\right) \ \chi^\a_{\l+a p\ {\rm mod}\ k}(\t,0).
\label{general.pf}
\ee
The $K_\l(\t,k\z;kp)$ are the characters for the
charge part, while the $\chi^\a_m(\t,0)$ are sums of $G/H$ characters 
for the neutral part, that are labelled
by the Abelian  number $m$ and other, possibly non-Abelian, quantum
numbers collectively denoted by $\a$. 
The explicit form of the neutral characters $\chi^\a_m$
is not needed, only their symmetries and modular transformations are 
relevant.

Equation (\ref{general.pf}) can be explained as follows.
The basic anyon has quantum numbers $(\l,m,\a)$, with
$m$ modulo $k$ and $\l$ modulo $kp$ owing to the periodicity:
\be
K_\l(\t,k\z;kp)=K_{\l+kp}(\t,k\z;kp)\ .
\label{pk-period}
\ee 
After adding one electron, the quantum numbers changes into
$(\l+p, m+p, \a)$; then, after adding $k$ electrons these numbers
return to those of the basic anyon. This explains the $k$ terms in the
sum (\ref{general.pf}).  The difference with respect to the $c=1$ case
(\ref{pf}) is that $n$-electron states couple to different neutral
parts for $n$ modulo $k$; actually, each $K_\l(\t,k\z;kp)$ in
(\ref{general.pf}) only sums electrons with $Q=\l/p +kn$, owing to its
different charge normalization.  Another way to state this fact is
that physical $\l$ and neutral $m$ charges are related by a $\Z_k$
parity rule \cite{cgt2}.

We can use pairs of these edge theories to model general interacting
topological insulators.  The functions $\Th_\l^\a(\t,\z)$ enjoy
similar properties under modular transformations as the $K_\l$ of
Section 3.1 and the partition function $Z^{NS}$, can be written
accordingly, that couple the up/down spin modes at one edge, though
the $U$ charge condition (\ref{U_transf}):
\be 
Z^{NS}=\sum_{\l,\a} \Th_\l^\a
\ov{\Th}_{-\l}^\a.
\label{NS.gen.th}
\ee
In this sum, the range for the $(\l,\a)$ values is given by the 
Wen topological order. In earlier works 
\cite{cz-mod}\cite{cgt1}\cite{cgt2}\cite{cv}, 
we showed that known expressions for
the partition functions of multicomponent Abelian theories in the
$K$-matrix formalism can be recast in the form of (\ref{general.pf}),
(\ref{NS.gen.th}); non-Abelian states are also written in this form,
as e.g. the Read-Rezayi parafermion states with $G/H=SU(2)/U(1)$.

Analogous expressions are obtained for the other spin sectors
$\wt{NS},R,\wt{R}$, by acting with the $T$ and $S$ transformations
as explained before: these partition functions take the same form 
(\ref{general.pf})
with signs for the $(-)^F$ weight and slightly different $\Z_k$
pairing in the Ramond sectors. 
Examples will be given later in this Section and in the Appendix.

\subsection{Stability argument}

The charge part $K_\l$ of the sectors $\Th_\l^\a$ in (\ref{general.pf}) is
parameterized by two independent numbers $(k,p)$, whose meaning can be
understood from the expression (\ref{pf}):

i) The values of the fractional charge are $Q=k\l/kp=\l/p$, $\l=0,\dots,p-1$,
and the minimal charge is equal to $e^*=1/p$.

ii) the Hall current (spin current) is obtained by applying
the $V$ transformation on (\ref{general.pf}), that acts on the charge
part $K_\l$, causing the shift of quantum numbers:
\be
V: \z  \to \z+\t, \qquad
\l \to\l +k, \qquad
\Delta Q^\uparrow =\nu^\uparrow =\frac{k}{p} ,
\label{k-hallcurr}
\ee
while the neutral characters in (\ref{general.pf}) are not affected.
We see that $\nu^\uparrow$ is parameterized by the ratio of the numbers 
$(k,p)$, but these might have common factors that are relevant 
in the stability argument.

As in Section 3, we should find the number of fluxes that creates an electron
excitation in the same anyon sector (fractional charge) and same neutral
sector (neutral quantum numbers), such that all fusion rules stay
unchanged \cite{ls}: owing to the periodicity of $K_\l$ in (\ref{pk-period}),
this number is given by $p$.
Then, the Fu-Kane-Mele flux argument considers the change in spin parity
due to adding $p/2$ fluxes. This is given by,
\be
V^{\frac{p}{2}}: \ \ \Delta S =
\Delta Q^\uparrow = 
{\frac{p}{2}}\ \nu^\uparrow = {\frac{k}{2}}.
 \label{phalf-fluxes}
\ee
Therefore, the Levin-Stern index (\ref{ls-index2}) for 
the spin parity anomaly is:
\be
2\Delta S= \frac{\nu^\uparrow}{e^*}=k.
\qquad
(-1)^{2\Delta S}=(-1)^{k}.
\label{ls-index-k}
\ee
 The stability analysis then continues by observing
that for odd values of $k$, the action of $V^{\frac{p}{2}}$ creates a
Kramers pair at the edge that is protected by TR symmetry; then, this
cannot be gapped and the topological insulator is stable.

The action on the anyon sectors (\ref{general.pf}) is,
\be
V^{\frac{p}{2}}: \ \ 
\Th_\l^\a(\t,\z) \ \to\  
\sum_{a=1}^k 
K_{\l+a p+kp/2}\left(\t, k\z; kp\right) 
\ \chi^\a_{\l+a p\ {\rm mod}\ k}(\t,0) \sim \Th_{\l'}^{\a'}(\t,\z), 
\label{act.pf}
\ee 
where the values of $(\l',\a')$ depends on the specific theory
considered through the symmetries of its characters.  Looking at the
expressions (\ref{general.pf}),(\ref{act.pf}), it is clear that the
neutral characters $\c^\a_m$ do not enter in the stability argument,
i.e. in the determination of the index (\ref{ls-index-k}); only the
charge parts are relevant.  Thus, the result (\ref{ls-index-k}) holds
for both Abelian and non-Abelian edge theories of topological
insulators.  In particular, the relevant parameters $(k,p)$ are
independent of the value of Wen topological order.
In conclusion, we have extended the Levin-Stern stability criterion
to any interacting topological insulator with time reversal symmetry.

In the case of multicomponent Abelian theories, the works
\cite{ls}\cite{chamon}\cite{vish} found the explicit edge interactions
that gap the system in the unstable cases, thus double checking the
result of the flux argument.  Such analysis of gapping interactions is
not yet available for general non-Abelian theories; however, we can
provide the following argument.  Some well-known non-Abelian states
have been described as projections of so-called ``parent'' Abelian
states \cite{cgt2}: for example, the $(331)$ state of distinguishable
electrons and its $k$-component generalizations are parents of the
Pfaffian and $\Z_k$ parafermion Hall states, respectively. The
non-Abelian theories are obtained by projecting the Abelian theories
to states that have identical electrons.  Since this projection does
not affect the TR invariance of states, it commutes with the analysis
of TR-invariant interactions in the Abelian theory and extends it to
these non-Abelian theories. In particular, we shall see later that the
topological insulator made by Pfaffian states is unstable.  Clearly,
the value of the Levin-Stern index is equal in the unprojected
(Abelian) and projected (non-Abelian) theories.


\subsubsection{Stability and modular non-invariance}

The stability of general topological insulators, corresponding to the
$\Z_2$ spin parity anomaly, is again accompanied by modular
non-invariance of the partition function. However, electromagnetic and
gravitational responses are not always equivalent as in the $c=1$ case
(neutral modes are clearly sensible to coordinate changes but not to
flux additions).

\begin{figure}[t]
\begin{center}
\includegraphics[width=15cm]{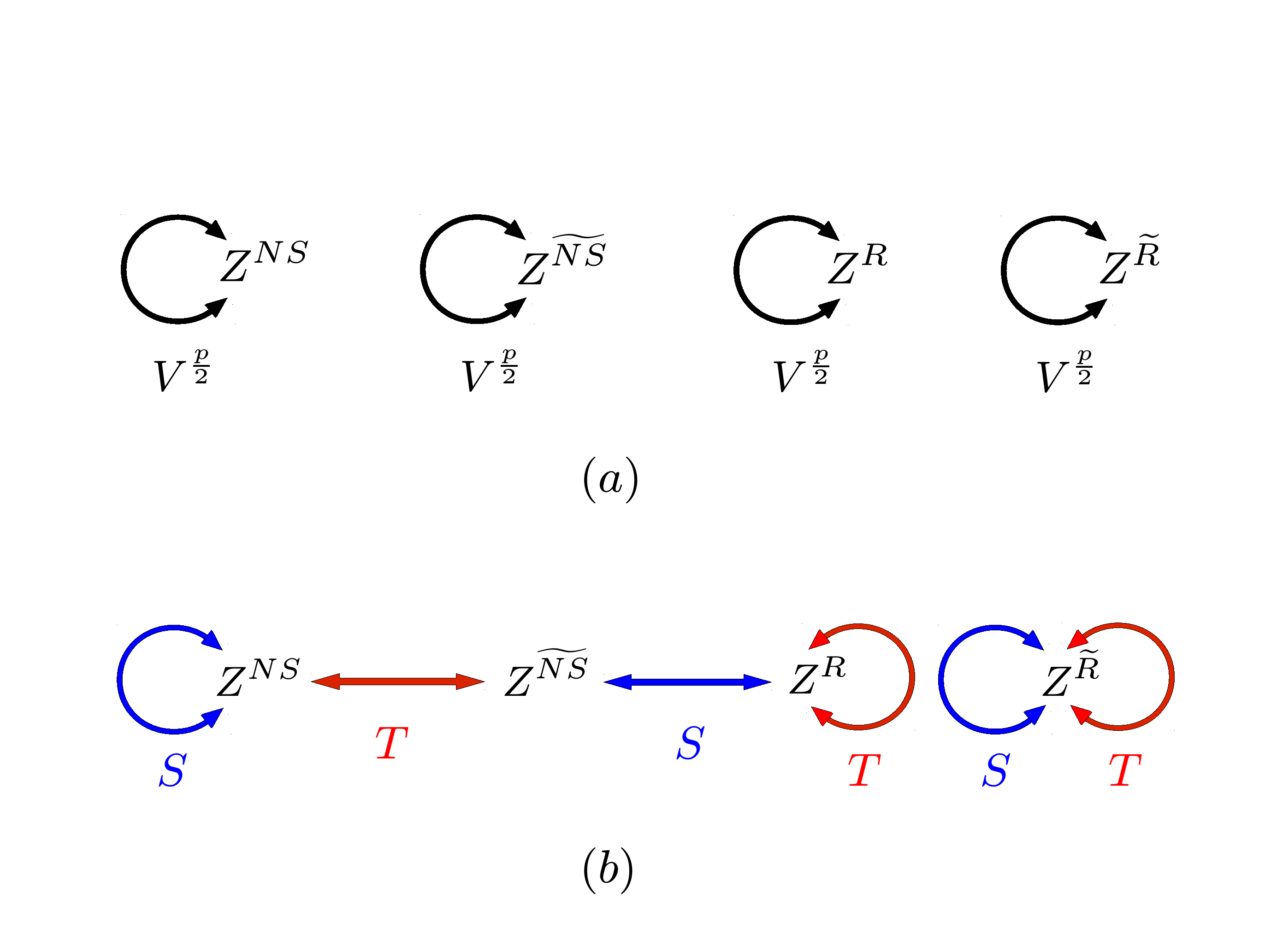}
\caption{Actions of (a) $p/2$ flux insertions and (b) 
modular transformations on the four spin sectors 
$Z^{NS}, Z^{\wt{NS}}, Z^R, Z^{\wt{R}}$ (even $p$ case).}
\label{fig7}
\end{center}
\end{figure}

We should distinguish the following cases, according to the parities
of $(k,p)$:

i) For $p$ odd, the action of $V^{\frac{p}{2}}$ is not a symmetry
of each spin sector and maps them one into another.
The transformations 
between Neveu-Schwarz and Ramond sectors and among their tildes
are the same as those of the $c=1$ theory (see Fig. \ref{fig6}(a) and 
Eq.(\ref{def_R})). 
The anyon sector $\Th_0^0$ containing the $NS$ ground state is naturally
mapped into $\Th_0^{R0}$ including the Ramond ground state.
The modular invariant and non-invariant partition functions are as in the
$c=1$ case:
\ba
Z_{\rm Ising}&=&Z^{NS} \! + Z^{\wt{NS}}\! +Z^R\! +Z^{\wt{R}},
\qquad\qquad\ \ k\ {\rm even,\ unstable},
\nl
 Z_{\rm TR}&=&\left(Z^{NS},Z^{\wt{NS}}, Z^R,Z^{\wt{R}} \right),
\qquad\qquad\quad  k\ {\rm odd,\ stable}.
\label{non-inv.pf}
\ea

ii) For $p$ even, the action of $V^{\frac{p}{2}}$ maps each spin sector into
itself and thus differs from the modular transformations
 (see Fig. \ref{fig7}).  For $k$ odd, the $Z_2$ anomaly manifests
itself within each spin sector, as a difference in spin parity between the
ground state and another ``anyon'' ground state (actually degenerate).
The TR symmetry of the theory then requires to splitting
each spin sector in two subsectors, 
$Z^\s\to (Z^\s_1,Z^\s_2)$, $\s=NS,\wt{NS},R,\wt{R}$, 
that are related by  $V^{\frac{p}{2}}$: 
$Z^\s_2=V^\frac{P}{2}\left(Z^\s_1\right)$ and collect anyon sectors
of same spin parity.
These subsectors carry a eight-dimensional representation of the 
modular group, namely the associated gravitational anomaly is 
slightly stronger.
Finally, for $k$ and $p$ both even, there is no anomaly and the
$Z_{\rm Ising}$ partition function is consistent with TR symmetry.
Summarizing, in all cases modular non-invariance is associated to stability
and the $\Z_2$ anomaly.

\subsection{Examples}

\subsubsection{Jain-like topological insulators}

The Jain states are the prominent multicomponent Abelian states
in the fractional quantum Hall effect. For $k$ components, their 
$c=k$ CFT involves a lattice of conformal dimensions 
(specified by the so-called $K$ matrix) in which
the Abelian symmetry $\wh{U(1)}^k$ is enhanced to 
$\wh{U(1)}\times \wh{SU(k)}_1$ \cite{wen-book}.
The fusion rules are Abelian but there are manifestations of the $SU(k)$
symmetry and its center $\Z_k$.

Consider the topological insulator made by a pair of spin up and spin down
Jain states; the filling fraction, minimal charge and stability index are:
\be
\nu^\uparrow=\frac{k}{2nk+1}, 
\qquad e^*=\frac{1}{2nk+1},
\qquad {2\Delta S = \frac{\nu^\uparrow}{e^*}=k},
\qquad (-1)^{2\D S}=(-1)^k,
\label{Jain-qn}
\ee
showing that the system is stable (unstable) for $k$ odd (even).

The Jain partition function in the $NS$ sector has been studied
extensively \cite{cz-mod}. The previous formulas identify the 
parameters entering the stability analysis are $k$ and $p=2nk+1$;
they have no common factor, $(k,p)=1$, and $p$ is always odd. 
Moreover, the topological order is equal to $p$. 
Thus, this case is very similar to that of Laughlin states,
and the pattern of flux insertions and modular transformations 
among the four spin sectors is the same as that already discussed
in Section 3 (Fig. \ref{fig6}). 
On the other hand, there are $(k-1)$ neutral modes
and the partition functions have
the simple-current form discussed in the previous Section.

The anyon sectors in the $NS$ sector take the form 
Eq.(\ref{general.pf}) \cite{cz-mod}:
\be
 \Theta^{NS}_\l(\tau,\zeta)=\sum_{a=0}^{k-1}
K_{\l+ap}\left(\t,k\z;kp\right)\  
\chi_{\l+ap\ {\rm mod}\ k}\left(\t,0\right),
\qquad\qquad \l=1,\dots,kp.
\label{JainNS-pf}
\ee
In this expression, $\chi_\b$ are the characters of the 
$\wh{SU(k)}_1$ affine algebra with Abelian label $\b$ modulo $k$ \cite{cft}. 
It is apparent that charged and neutral sectors are paired by
the $\Z_k$ section rule $\l=\b$ modulo $k$, whose
origin can be understood as follows \cite{cz-mod}\cite{cgt1}.
The Abelian $k$-dimensional lattice of conformal dimensions specified by the
$K$ Gram matrix is not orthogonal, i.e.
the charged and neutral excitations are not independently
generated. The decomposition underlying the expression 
(\ref{JainNS-pf}) is obtained by embedding the  Abelian
lattice into a thinner one that is orthogonal in the 
charge and neutral directions. In this last lattice, 
the ``physical'' points are those obeying the $\Z_k$ selection rule.

The anyon sectors are periodic by $\Theta^{NS}_{\l+p}=\Theta^{NS}_\l$,
due to $p=1$ modulo $k$; thus there are $p$ independent ones
(the value of the topological order), that can be chosen to be
$ \Theta^{NS}_{ka}$, $a=1,\dots,p$.
Let us present more explicit formulas in the $k=2$ case (unstable);
the partition functions of the four spin sectors are:
\ba
\label{Jain-totpf}
\Theta^{NS}_{2a}(\tau,\zeta) &=& 
K_{2a} ~\chi_{0} + K_{2a+p}~ \chi_{1},\qquad a=1,\dots,p,
\nl
\Theta^{\widetilde{NS}}_{2a}(\tau,\zeta) &=& 
K_{2a} ~\chi_{0} - K_{2a+p}~ \chi_{1},
\nl
\Theta^{R}_{2a}(\tau,\zeta) &=& 
K_{2a} ~\chi_1 + K_{2a+p}~ \chi_0,
\nl
\Theta^{\widetilde{R}}_{2a}(\tau,\zeta)  &=& 
-K_{2a} ~\chi_1 + K_{2a+p}~ \chi_0,
\label{jain2-pf}
\ea
where $K_\l(\t,2\z;2p)=K_{\l+2p}(\t,2\z;2p)$ and $\c_\s=\c_{\s+2}$.
Note that first (second) $K_\l$ in each
expression includes an odd (even) number of fermion excitations;
then, the relative sign appearing in the $\widetilde{NS}$ and $\widetilde{R}$ 
expressions accounts for the $(-1)^F$ factor in the trace (\ref{pf2-trace}).
Note also that the $\Z_2$ parity rule between charge and neutral
quantum numbers that is different in the $NS$ and $R$ sectors. 
The transformation properties of these sectors are given in Appendix A.
The partition functions of the four sectors are:
\be
Z^{(\s)}=\sum_{a=1}^p \Th^{(\s)}_{2a} \ov{\Th^{(\s)}_{-2a}},
\qquad\qquad 
\s=NS,\wt{NS},R,\wt{R}.
\label{jain2-totpf}
\ee
The partition function for this unstable system is given by $Z_{\rm Ising}$.


\subsubsection{Multicomponent Abelian topological insulators}

We briefly discuss the general Abelian edge theories that have been
studied extensively in \cite{ls}\cite{chamon}\cite{vish} and
show how they fit in the present analysis.
Abelian conformal theories with central charge $c=n$ are characterized by 
a $2n$-dimensional lattice of conformal weights and charges, that
specifies the statistical phases $\theta/\pi$, the electric charge
$Q$ and spin $S$ of excitations through the following formulae,
\be
\frac{\th}{\pi}=\sum_{i,j=1}^{2n}{\bf n}_i{\bf K}^{-1}_{ij}{\bf n}_j,
\qquad
Q=\sum_{i,j=1}^{2n}{\bf t}_i{\bf K}^{-1}_{ij}{\bf n}_j,
\qquad
S=\sum_{i,j=1}^{2n}{\bf s}_i{\bf K}^{-1}_{ij}{\bf n}_j,
\qquad {\bf n}_i \in \Z^{2n}.
\label{ab-spec}
\ee
These involve the Gram matrix ${\bf K}^{-1}$ of the lattice,
the charge and spin vectors, ${\bf t}$, ${\bf s}$, 
and the integer vector ${\bf n}_i$ specifying the excitations.
In particular, excitations with integer statistics, i.e. electrons
and their compounds, are described by the dual lattice of ${\bf K}$.
The Gram matrix is expressed by the basis vectors ${\bf v}_i$ of the lattice
as ${\bf K}^{-1}_{ij}= {\bf v}_i\cdot {\bf \eta}\cdot {\bf v}_j$,
where ${\cal \eta}_{ij}=\d_{ij}\s_i$, $\s_i=\pm 1$ is the signature
of the Lorentzian metric, expressing the chirality of
excitations \cite{ctz}\cite{wen-book}.
Basis vectors $({\bf v}_i)_\a$ and
rescaled eigenvectors $\sqrt{\l_\a}({\bf u}_\a)_i$ form equal matrices,
up to transposition.

In time reversal invariant systems, 
there is an equal number of positive and negative chiralities.
Next, we can choose a basis in which the first (second) 
$n$ components describe the spin-up (down) modes.
Then the action of the TR transformation ${\bf T}$ and the
other quantities take a two-dimensional block form:
\be
{\bf T}  =\left( \begin{array}{cc}
0 & 1 \\ 1 & 0
\end{array} \right),
\qquad
{\bf t}  =\left( \begin{array}{c}
t \\ t \end{array} \right),
\qquad
{\bf s}  =\left( \begin{array}{c}
s \\ -s
\end{array} \right),
\qquad
{\bf n}  =\left( \begin{array}{c}
n^\uparrow \\ n^\downarrow
\end{array} \right).
\label{two-block}
\ee
The TR symmetry of the spectrum implies
the condition ${\bf K= - T\, K\, T}$, with solution:
\be
{\bf K} =\left(
\begin{array}{cc}
K & W \\ W^T & -K
\end{array}
\right),
\qquad \quad K^T=K, \ \ \ W^T=-W.
\label{gen-K}
\ee
It follows that ${\bf K}$ has eigenvectors in pairs $(\l_\a,-\l_\a)$,
that are integer valued \cite{ls}\cite{chamon}\cite{vish}.

The case of Jain states considered before corresponds to $W=0$,
and, moreover, to a positive definite metric $K$, because
excitations of same spin have same chirality:
\be
{\bf K} =\left(
\begin{array}{cc}
K & 0 \\ 0 & -K
\end{array}
\right),
\qquad
n^\uparrow_i K_{ij} n^\uparrow_j \ge 0.
\label{block-K}
\ee
The partition functions of topological insulators made of these
Abelian theories can be obtained within the $K$-matrix formalism, and
the expressions for anyon and spin sectors are given in Appendix A.
The decomposition of anyon sectors in simple-current form
(\ref{general.pf}), schematically 
$\Th^\uparrow=\sum\, K^\uparrow\, \c^\uparrow$, is obtained 
by separating the lattice into charged and neutral sublattices,
whose points are coupled by a parity rule as explained before.

Next we discuss the systems with neutral modes counter-propagating
with respect to the charged mode of the same spin type. These are
described by the block-diagonal form (\ref{block-K}), where $K$ has
mixed signature, i.e. is not positive definite: an example is given by
the Jain states with $\nu^\uparrow=k/(2nk-1)$. As discussed in
\cite{cz-mod}, the simple current expression (\ref{general.pf}) of
partition functions is still valid with the replacement $\c_a \to
\ov{\c}_a$ for the opposite neutral chirality, schematically
$\Th^\uparrow=\sum\, K^\uparrow\, \ov{\c}^\uparrow$.

Finally, the case $W\neq 0$ describes interacting excitations with
up and down spins.  Indeed, let us choose the standard basis
specified by ${\bf t}=(1,1)$ and ${\bf s}=(1,-1)$; the statistical phases 
are  computed from $\th/\pi=\ell_i {\bf K}_{ij} \ell'_j$,
where the $i$-th electron is represented by 
$\ell_j=e(i)_j=\d_{ij}$. Then, the correlation
of up and down electrons, 
$\langle\Psi_j^\uparrow(x)\Psi_k^\downarrow(y)\rangle$,
has a phase proportional to  $W_{jk}\neq 0$, for
$1\le j \le n$ and $n<k\le 2n$.
This is due to the coupling of the charge mode, e.g. spin up,  
with neutral modes of opposite spin. In this case, 
the simple current decomposition
of anyon sectors (\ref{general.pf}) applies in the form  
$\Th^\uparrow=\sum\, K^\uparrow\, \c^\downarrow$.
In conclusion, the simple current form of the partition function is valid
in all Abelian cases; the stability argument discussed earlier
applies equally, because it does not involve the properties of 
neutral modes.

Let us finish the discussion of Abelian theories by considering the 
example of type (\ref{block-K}) with two-component $K$ matrix,
\be
K= \left(
\begin{array}{cc}
3 &1 \\ 1 &5
\end{array}
\right),
\label{K-3115}
\ee
and  charge vector $t=(1,1)$. Using previous formulas,
we obtain the charge spectrum, minimal charge and filling fraction
of the corresponding Hall state:
\be
Q=\frac{2n_1+n_2}{7}, \quad n_1,n_2\in \Z, \qquad e^*=\frac{1}{7},\qquad 
\nu^\uparrow=\frac{3}{7},
\label{qn-3115}
\ee
while the topological order is $\det K=14$.

Next we consider the topological insulator made by a pair of these chiral
states.  The parameters entering the stability analysis are $(k,p)=(3,7)$,
thus the topological state is stable, 
\be 
2\Delta S=\frac{\nu^\uparrow}{e^*}=3, \qquad (-1)^{2\D S}=-1.
\label{two-comp}
\ee
This example shows that the stability is independent of
the number of fermion modes and of the value of the topological order.


\subsubsection{(331) and Pfaffian topological states}

The Pfaffian state is the simplest  example of non-Abelian
quantum Hall states \cite{mr}. Following the previous 
discussion of stability, it can be
analyzed together with its parent Abelian state, the so-called
$(331)$ which has the same
charge spectrum and filling fraction (see Ref. \cite{cgt2} for a
complete analysis of this relation). 
From the matrix $K=\left((3,1),(1,3)\right)$, we find:
\be 
\nu^\uparrow=\frac{1}{2},\qquad
e^*=\frac{1}{4},\qquad  2\Delta S= 2, 
\qquad (-1)^{2\Delta S}=1.
\label{qn-331}
\ee
Thus the topological insulators made by pairs of these Hall states 
are unstable. The parameters entering the stability 
analysis are $(k,p)=(2,4)$. 

Let us first discuss the Abelian state. Its topological order is
$\det{K}=8$, thus there are $8$ anyon sectors and a $\Z_2$ parity rule.
In the $NS$ sector, they read:
\be
\Th_\l=K_\l\,\c_\l+K_{\l+4}\,\c_{\l+2},\qquad\quad \l=1,\dots,8,
\label{331-NSanyon}
\ee
where $K_\l=K_\l(\t,2\z;8)$ and $\c_l=K_\l(\z,0;4)$.
The Neveu-Schwarz partition function reads \cite{cgt1}:
\be
Z^{NS}_{(331)}=\sum_{a=0}^3
\left\vert K_a\,\c_a + K_{a+4}\, \c_{a+2} \right\vert^2+
\left\vert K_a\,\c_{a+2} + K_{a+4}\, \c_a \right\vert^2.
\label{331-NSpf}
\ee
Note that the four charge sectors appears twice in the spectrum
coupled to different neutral parts. The expressions for the other
spin sectors are given in Appendix.

In the Pfaffian state, the neutral part is provided by the characters
of the Ising model, i.e. the $Z_2$ parafermions, $\c_a^\ell$, labelled by
$a$ modulo $4$ and $\ell=0,1,2$, non-vanishing for 
$a=\ell$ modulo $2$, and obeying $\c_{a+2}^\ell=\c_2^{2-\ell}$ \cite{cv}.
There are only three independent values, $\c_0^0=\c_2^2=I$, 
$\c_1^1=\c_3^1=\s$ and $\c_2^0=\c_0^2=\psi$, denoted as the corresponding
conformal fields.
The anyon sectors are ($NS$ sector):
\be
\wt{\Th}_a^\ell=K_a\,\c_a^\ell+K_{a+4}\,\c_{a+2}^\ell,
\qquad\quad a=0,1,2,3,\quad
\ell=0,1,2,\quad a=\ell\ {\rm mod}\ 2.
\label{pfaffian-anyon}
\ee
It is apparent that the charge parts in the $(331)$ and Pfaffian 
states are equal, while the neutral parts of the latter obey some 
symmetries that reduce the
topological order from $8$ to $6$; one indeed checks that
$\Th^1_1=\Th^1_5$ and $\Th^1_3=\Th^1_7$ \cite{cgt1}.
The $NS$ partition function reads \cite{cv}:
\ba
Z^{NS}_{\rm Pf}&=&\sum_{a=0,2}
\left\vert K_a\,\c^0_a + K_{a+4}\, \c^0_{a+2} \right\vert^2+
\left\vert K_a\,\c^0_{a+2} + K_{a+4}\, \c^0_a \right\vert^2+
\nl
&&\qquad \left\vert \left(K_1+K_{-3} \right)\c_1^1 \right\vert^2+
\left\vert \left(K_3+K_{-1} \right)\c_1^1 \right\vert^2,
\label{pfaffian-NS}
\ea
where charge sectors appear again twice, but are coupled differently
to neutral states. In particular, the two $(331)$ sectors with neutral
charge $a=1$ (resp. $a=-1$) are projected to a single one.  This
example clarifies that the Fu-Kane-Mele stability argument is the same
in both theories, since it only deals with the charge parts.  The
expressions of the other spin sectors for the Pfaffian topological
state are also given in the Appendix.

These unstable theories are characterized by $(k,p)$ both even, thus
flux insertions and modular transformations act differently on the
four spin sectors (Fig. \ref{fig7}).  As discussed before,
$V^{\frac{p}{2}}=V^2$ is an isometry of each spin sector.  Since there
is no associated $\Z_2$ anomaly, the modular invariant $Z_{\rm Ising}$
partition function is consistent with TR symmetry.


\subsubsection{Read-Rezayi parafermionic states}

The Read-Rezayi states \cite{rr} are generalization of the Pfaffian
state involving neutral modes of the $\Z_k$ parafermions that can be
described by the coset $\wh{SU(2)}_k/U(1)$ \cite{cv}.  The quantities
entering in the stability index (\ref{ls-index}) are
\be
\nu^\uparrow=\frac{k}{kM+2}, 
\qquad e^*=\frac{1}{kM+2},
\qquad  2\Delta S =k, \qquad
(-1)^{2\D S}=(-1)^k,
\label{rr-qn}
\ee
where $k=3,4\dots$ and $M=1,3,\dots$. In this case, $(k,p)=(k,kM+2)$, thus
the topological insulators made by a pair of these states is stable
(unstable) for $k$ odd (even). Note that $k$ and $p$ have the same parity:

i) For $k$ and $p$ odd, these numbers have no common factor, and the flux
insertions and modular transformations follow the pattern of the 
stable, odd $k$ Jain states and of the $c=1$ theory (Fig. \ref{fig6}).
The modular non-invariant partition function takes the form $Z_{\rm TR}$
in (\ref{non-inv.pf}).

ii) For  $k$ and $p$ even, they have a common factor of $2$ and the
flux insertions and modular transformations are the same as those
of the $k=2$ case discussed in the previous paragraph (Fig. \ref{fig7}). 
The modular invariant partition function is $Z_{\rm Ising}$.

All partition functions and modular transformations
are described in Appendix. Let us briefly discuss the form of
the Neveu-Schwarz anyon sectors, taken from Ref.\cite{cv}.
These read:
\be
\Th^\ell_a = \sum_{b=1}^k K_{a+bp}\left(\t,k\z;kp\right) 
\c^\ell_{a+2b}(\t,0,2k), \qquad a=\ell\ \ {\rm mod} \ 2,\quad
p=2+kM.
\label{rr-NSanyon}
\ee
The charge characters $K_\l$ with periodicity $kp$, are coupled
to the $\Z_k$ parafermion characters $\c^\ell_m$, that are specified by
the $SU(2)_k$ quantum number $\ell=0,1,\dots,k$, and the Abelian number
$m$ modulo $2k$. The $\Z_k$ parity rule between the two Abelian numbers
is $\l=m$ modulo $k$ (note that $p=2$ modulo $k$).
The parafermion characters obey the periodicities
$\c^\ell_m=\c^\ell_{m+2k}=\c^{k-\ell}_{m+k}$ and vanish for
$m+\ell=1$ modulo $2$. Taking into account these properties, one finds
the periodicity $\Th^\ell_{a+p}=\Th^{k-\ell}_a$, implying $p(k+1)/2$
independent anyon sectors, the topological order.

\section{Conclusions}

In this paper we have obtained the general form of the partition
function of edge excitations in non-chiral topological states
protected by time-reversal symmetry, such as the quantum spin Hall
effect and topological insulators.  The study of partition functions
has allowed us to discuss the flux argument for the stability of the
topological phase in great generality and to extend it to interacting
systems possessing non-Abelian edge excitations.

We have emphasized the anomaly in the $\Z_2$ edge spin parity that is
associated to stable topological phases.  We considered the modular
transformations of the partition function that map the four spin
sectors, Neveu-Schwarz, Ramond and their tildes, among themselves.  We
have found that the $\Z_2$ anomaly is accompanied by modular
non-invariance, that is a kind of discrete gravitational anomaly.  We
also discussed the cases in which the electromagnetic and
gravitational responses are equivalent and when they are different.

Among the possible directions of future progress, we mention the
extension of our stability analysis to topological superconductors
\cite{qz-rev}.  Let us briefly discuss this point.  From the point of
view of edge dynamics, topological superconductors amount to systems
of (interacting) $N_f$ neutral chiral Majorana modes with spin up and
$N_f$ ones with opposite spin and chirality, thus making $N_f$ Ising
CFTs \cite{cft}.  The topological phases are protected by the
$\Z_2\times\Z_2$ symmetry of independent edge spin parity for each
chirality, $(-1)^{2S^\uparrow}=(-1)^{N_\uparrow}$ and 
$(-1)^{2S^\downarrow}=(-1)^{N_\downarrow}$. This
follows from TR symmetry and the absence of spin-flip terms in the
bulk Hamiltonian \cite{qz-rev}.  Since this symmetry does not allow
any mass term, the non-interacting topological insulators are
classified by the $\Z$ index equal to $N_f$.

On the other hand, for $N_f=8$ a non-trivial quartic interaction was found
that gaps the system without breaking the symmetry explicitly or 
spontaneously \cite{fk}\cite{qi}.
This interaction  is possible because the Ising CFTs have spin
$\s$ and disorder $\mu$ fields of dimensions $h=1/16$, such that the
product of eight of them has dimension
$h=1/2$ and can bosonized and refermionized to obtain a
quartic fermionic term respecting the symmetry.
Therefore, the classification of interacting topological superconductors
is associated to a $Z_8$ index (at least).

For even $N_f$, the free Majorana fermions have $c=N_f/2$ and 
yield the same fermion systems considered in this
paper. However, the neutral modes cannot be probed by
flux insertions for studying stability.
In an interesting paper \cite{rz}, Ryu and Zhang suggested to consider the 
modular non-invariance as a test of stability of topological superconductors.
In practice, they associated the instability to the modular invariance
of the partition function for the singlet sector of the theory
with respect to the $\Z_2\times\Z_2$ symmetry, 
$(-1)^{N_\uparrow}=(-1)^{N_\downarrow}=1$, i.e. involving only bosonic
chiral states.
Indeed, this projected partition function is modular invariant for $N_f=8$, 
as it corresponds to the so-called Gliozzi-Scherk-Olive projection 
$Z_{\rm GSO}$ in  superstring theory \cite{gso}. 
This results provided a symmetry argument for the instability 
of $N_f=8$ topological superconductors \cite{fk}\cite{qi} (also extended to
other systems in \cite{ryu}). 

In our analysis, we considered the partition function of the full theory
$Z_{\rm Ising}$ that is always modular invariant, but argued that it
might not be consistent with TR invariance, owing to anomalies 
in some of its sectors.
The two criteria of stability do not seem to be equivalent in general,
but at least they agree on the instability of the $N_f=8$ case.
The argument goes as follows. 
In Section 3, we have seen that in free fermionic systems, the half flux
insertion maps Neveu-Schwarz into Ramond sectors, i.e. we have the equivalence
$V^{\frac{1}{2}}\sim ST$, between electromagnetic and gravitational responses.
Moreover, $V^{\frac{1}{2}}$ does the correct flux insertion for 
measuring the anomaly.  We then consider the chiral spin of the excitation
in the $R$ sector obtained from the $NS$ sector by the $ST$
transformation: for $N_f=2N$ fermion modes, this is $\Delta S^\uparrow=\Delta
S^\downarrow =N/4$. The first $\Z_2\times\Z_2$ anomaly free case is 
for $N_f=8$, with chiral indices:
\be
\qquad\quad (-1)^{N_\uparrow}=(-1)^{N_\downarrow}=1,\qquad\quad N_f=8.
\label{8fermions}
\ee
Therefore, the modular invariant partition function $Z_{\rm Ising}$
is free of $\Z_2\times\Z_2$ anomalies for $N_f=8$, thus confirming the
instability of this system in our analysis.

More general results cannot be provided at present, owing to two main
difficulties: the correct measure of spin anomaly in other systems and
the form of the partition function to be associated to the edge of interacting
topological superconductors. 
Regarding the second point, the absence of the $U$ condition, 
matching the charges of the two chiralities, 
allows many possible expressions for the edge partition function,
of the form $Z=\sum {\cal N}_{\l\mu} K^\uparrow_\l
\ov{K}^\downarrow_\mu$, with ${\cal N}$ any symmetric positive integer matrix
commuting with the $S$ and $T$ transformations \cite{cft}.

Another possible development concerns the comparison with the stability
criterion proposed in \cite{levin}, which is based 
on properties of fusion rules,
again related to modular transformations through the Verlinde formula.
Finally, there is the study of stability of three dimensional systems:
in particular, recent works  \cite{3d-interact} suggest the existence of
topological states at the $2d$ surface of a $3d$ topological
insulators, whose stability could be different from that
of strictly $2d$ systems considered here.

{\bf Acknowledgments}

AC and ER thanks T. H. Hansson, D. Seminara, G. Viola, P. Wiegmann and 
G. R. Zemba for interesting
discussions. This work was partially supported by the EC grant
M. Curie Action IRSES-295234 ``Quantum Integrability, Conformal Field 
Theory and Topological Quantum Computation''.

\appendix

\section{Modular transformation and spin structures}

In the following we give the expressions of partition functions for
the four spin sectors $NS,\ \wt{NS},\ R,\ \wt{R}$, that describe the
edge excitations of topological insulator models examined in the main
text. We describe their behavior under
flux insertion and modular transformations.

\subsection{Laughlin states}
\label{app.c1}

The Laughlin states discussed in Section 3 are described by the $c=1$
CFT \cite{cft}. The anyon sectors for the chiral modes of
the $NS$ and $\wt{NS}$ spin sectors are, for 
$\lambda=1,\cdots,p$, $p$ odd, \cite{cz-mod}:
\begin{align}
\label{NSchar1}
&K^{NS}_\l(\t,\z;p)=\frac{F(\t,\z)}{\eta(\tau)}
\sum_{n\in\Z} \exp\left(i2\pi\left( \t \frac{(np+\l)^2}{2p} + 
\z \frac{np+\l}{p}\right) \right),\\
&K^{\widetilde{NS}}_\l(\t,\z;p)=\frac{F(\t,\z)}{\eta(\tau)}
\sum_{n\in\Z}(-1)^{pn} \exp\left(i2\pi\left( \t \frac{(np+\l)^2}{2p} + 
\z \frac{np+\l}{p} +\frac{\l}{2} \right)\right),\notag
\end{align}
with $F=\exp\left[-\pi (\I \z)^2/p\,\I \t \right]$ is a non-holomorphic
prefactor and $\eta(\t)$ the Dedekind function 
\be
\label{dedekind}
\eta(\tau)=q^{\frac{1}{24}}\prod_{n=1}^{\infty}(1-q^n),
~~~~~q=\text{exp}(i2\pi\t).  
\ee 
The anyon sectors of the $R$ and
$\widetilde{R}$ spin sectors are defined by:
\begin{align}
&K^{ R}_{\lambda}= K^{NS}_{\lambda+\frac{p}{2}}, \notag\\ 
&K^{ \widetilde{R}}_{\lambda}= K^{  \widetilde{NS}}_{\lambda+\frac{p}{2}} .
\label{Rchar1}
\end{align}
The edge partition functions for each spin sector (Eq.\eqref{pf-trace}
and \eqref{pf2-trace}) are obtained by matching the fractional charge
of the chiral and antichiral anyon sectors locally at the edge (see
$U$ condition (\ref{U_transf})),
\begin{equation}
\label{pfc1-spinsector}
Z^{(\s)}=\sum_{\lambda=1}^p K^{ (\s)}_{\lambda} \ov{{K}^{
  (\s)}_{-\lambda}},~~~~~ \s=NS, \widetilde{NS}, R, \widetilde{R}.
\end{equation}

The transformation of the anyon sectors ((\ref{NSchar1}) and
(\ref{Rchar1})) under the modular group, generated by $S$ and $T$, and
for the insertion of one and $p/2$ fluxes through the annulus, the $V$
and $V^{\frac{p}{2}}$ transformations, respectively,
are obtained by extending the
calculations of Ref. \cite{cz-mod} \cite{cv}. Altogether
they are:
\begin{itemize}
\item $S$
  \begin{align}
\label{S.transf1}
    &K^{ NS}_{\lambda} (-\frac{1}{\tau},\frac{-\zeta}{\tau}) =
e^{i\varphi}\sum_{\lambda^{\prime}=1}^{p}S_{\lambda\lambda^{\prime}}~ 
K^{NS}_{\lambda^{\prime}}(\tau,\zeta),   \\
    &K^{ \widetilde{NS}}_{\lambda} (-\frac{1}{\tau},\frac{-\zeta}{\tau})= 
e^{i\varphi}\sum_{\lambda^{\prime}=1}^{p}S_{\lambda\lambda^{\prime}}
K^{R}_{\lambda^{\prime}}(\tau,\zeta),\notag\\
    &K^{ R}_{\lambda} (-\frac{1}{\tau},\frac{-\zeta}{\tau}) =
e^{i\varphi} \sum_{\lambda^{\prime}=1}^{p}S_{\lambda\lambda^{\prime}}~ 
K^{\widetilde{NS}}_{\lambda^{\prime}}(\tau,\zeta),\notag\\
    &K^{ \widetilde{R}}_{\lambda} (-\frac{1}{\tau},\frac{-\zeta}{\tau}) = 
\text{exp}\bigg( 2 \pi i \frac{p}{4}\bigg)~e^{i\varphi}
\sum_{\lambda^{\prime}=1}^{p}S_{\lambda\lambda^{\prime}} 
K^{ \widetilde{R}}_{\lambda^{\prime}}(\tau,\zeta),\notag 
\end{align}
with 
\begin{equation}
\label{S-element}
S_{\lambda\lambda^{\prime}}= \frac{1}{\sqrt{p}}  
\bigg( 2\pi i \frac{\lambda \lambda^{\prime} }{p} \bigg), 
~~~~~~~e^{i\varphi} =
\text{exp}\bigg( \frac{i\pi}{ p} \text{Re}
\bigg(\frac{\zeta^2}{\tau}\bigg)\bigg). 
\end{equation}
\item $T$
\begin{align}
\label{T-transf1}
    &K^{NS}_{\lambda}(\tau+1, \zeta)= \text{exp}
\bigg( -2\pi i \frac{\lambda}{2}\bigg)T_a
K^{ \widetilde{NS}}_{\lambda}(\tau,\zeta),   \\
    &K^{\widetilde{NS}}_{\lambda}(\tau+1, \zeta)= 
\text{exp}\bigg( 2\pi i \frac{\lambda}{2}\bigg)T_a 
K^{NS}_{\lambda}(\tau, \zeta),\notag\\
    &K^{ R}_{\lambda}(\tau+1, \zeta)= T_aT_b
K^{R}_{\lambda}(\tau, \zeta),\notag\\
    &K^{ \widetilde{R}}_{\lambda} (\tau+1, \zeta)= 
 T_aT_bK^{ \widetilde{R}}_{\lambda}(\tau, \zeta),\notag 
\end{align}
with
\be
\label{T-element}
T_a= \text{exp} \bigg( 2 \pi i \bigg( \frac{\lambda^2}{2p} -
\frac{1}{24} \bigg)\bigg),
~~~~T_b= \text{exp}\bigg( 2 \pi i \bigg( \frac{p}{8} +
\frac{\lambda}{2} \bigg)\bigg).
\ee
\item $V$
\begin{align}
\label{V-transf1}
    &K^{ NS}_{\lambda}(\tau, \zeta+\tau)= 
V_{\Phi_0}K^{NS}_{\lambda+1}(\tau,\zeta),      \\
     &K^{ \widetilde{NS}}_{\lambda}(\tau, \zeta+\tau)= 
-V_{\Phi_0}K^{\widetilde{NS}}_{\lambda+1}(\tau,\zeta),\notag\\
      &K^{ R}_{\lambda}(\tau, \zeta+\tau)= 
V_{\Phi_0}K^{ R}_{\lambda+1}(\tau,\zeta),\notag\\
    &K^{ \widetilde{R}}_{\lambda}(\tau, \zeta+\tau)= 
-V_{\Phi_0}K^{ \widetilde{R}}_{\lambda+1}(\tau,\zeta),\notag
\end{align}
with
\be
\label{V-element}
V_{\Phi_0}(\tau,\zeta)= \text{exp} \bigg( -2 \pi i \frac{1}{p} 
\bigg( Re\frac{\tau}{2} +Re \zeta \bigg)  \bigg)
\ee
\item  $V^{\frac{p}{2}}$
\begin{align}
\label{Vp2-transf1}
    &K^{ NS}_{\lambda} (\tau,\zeta +  \frac{p \t}{2}) =
V_{\frac{p}{2}\Phi_0} ~K^{ R}_{\lambda},  \\
    &K^{ \widetilde{NS}}_{\lambda}(\tau,\zeta +  \frac{p \t}{2}) =
V_{\frac{p}{2}\Phi_0} \text{exp}\bigg(-2\pi i \frac{p}{4} \bigg) 
K^{\widetilde{R}}_{\lambda},\notag\\
     & K^{ R}_{\lambda} (\tau,\zeta +  \frac{p \t}{2}) =
V_{\frac{p}{2}\Phi_0}K^{ NS}_{\lambda},\notag\\
    &K^{ \widetilde{R}}_{\lambda}(\tau,\zeta +  \frac{p \t}{2}) =
V_{\frac{p}{2}\Phi_0} \text{exp} \bigg(-2\pi i \frac{p}{4} \bigg) 
K^{ \widetilde{NS}}_{\lambda},\notag 
\end{align}
with
\be
\label{Vp2-element}
V_{\frac{p}{2}\Phi_0}=  \text{exp}\bigg(-2 \pi i \bigg( \frac{p}{8}Re\tau+
\frac{1}{2}Re\zeta \bigg) \bigg).
\ee
\end{itemize}
Upon using these formulas, we obtain that the transformations of
partition functions (\ref{pfc1-spinsector}) 
illustrated in Fig.(\ref{fig6}). As discussed
in Section 3.4, the modular invariant partition function
$Z_{\text{Ising}}$ (\ref{Ising}) is not consistent with TR symmetry
owing the presence of the $Z_2$ anomaly, then the Laughlin-type
topological insulators are stable.


\subsection{Jain states}
\label{app.jain}
The topological insulators with two-component
Jain edge systems discussed in Section
4.3.1, are described by the $c=2$ CFT with symmetry
$\widehat{U(1)}\times \widehat{SU(2)}_1$; the two parameters that
enter the stability argument (see Section 4.2) are
$(k,p)=(2,4n+1)$. The characters of the $\widehat{U(1)}$ charge part
are expressed by the functions $K_{2a+\alpha p}(\tau,2\zeta; 2p)$
of the previous section (Eq.(\ref{pf})), where $\alpha=0,1$ and
$a=1,\cdots,p$. The $\widehat{SU(2)}_1$ neutral characters are:
\begin{equation}
\label{Jain-neutral}
\chi_{\alpha=0} =\frac{1}{\eta(q)} \sum_{n \in \Z}q^{n^2},
~~~~~\chi_{\alpha=1} =\frac{1}{\eta(q)} \sum_{n \in\Z}q^{(2n+1)^2 /4}.
\end{equation}
These have the periodicity $\c_{\a+2}(\t,0)=\c_{\a}(\t,0)$ and obey
the modular transformations (with $k=2$) \cite{cz-mod}
\begin{align}
\label{mod-neutro-jain}
    &T:~~~\c_{\a}(\t+1,0)=\text{exp}\bigg(2 \p i\bigg(\frac{\a(k-\a)}{2k}-
\frac{(k-1)}{24} \bigg)\bigg) \c _{\a}(\t,0),   \\
    &  S:~~~\c_{\a}(-\frac{1}{\t},0)=
\frac{1}{\sqrt{k}}\sum_{\a'=1}^{k}\text{exp}
\bigg(-2\p i\frac{\a\a'}{k} \bigg)\c_{\a'}(\t,0).\notag
\end{align}

Then,  the chiral anyon sectors for the four spin sectors are:
\begin{align}
\label{jain-anyon2}
&\Theta^{NS}_{2a}(\tau,\zeta) = K_{2a} ~\chi_{0} + K_{2a+p}~
\chi_{1},\\ &\Theta^{\widetilde{NS}}_{2a}(\tau,\zeta) = K_{2a}
~\chi_{0} - K_{2a+p}~\chi_{1},\notag\\ &\Theta^{R}_{2a}(\tau,\zeta) =
K_{2a}~\chi_{1}+ K_{2a+p} ~\chi_{0},\notag\\ &\Theta^{
  \widetilde{R}}_{2a}(\tau,\zeta) = - K_{2a} ~\chi_{1} + K_{2a+p}
~\chi_{0}\notag.
\end{align}
For each spin sectors the edge partition functions  are obtained
by coupling the chiral and antichiral modes and summing over
$a=1,\cdots, p$, as follows (see Eq.(\ref{jain2-totpf})):
\be
Z^{(\s)}=\sum_{a=1}^p \Th^{(\s)}_{2a} \ov{\Th}^{(\s)}_{-2a},
\qquad\qquad 
\s=NS,\wt{NS},R,\wt{R}.
\label{jain-pf-tot}
\ee

Under the modular group and the insertion of fluxes, the anyon sectors
(\ref{jain-anyon2}) possess the following transformation properties:
\begin{itemize}
\item $S$
\begin{align}
\label{Sjain-tranf}
& \Theta^{ NS}_{2a} (-\frac{1}{\tau},-\frac{\zeta}{\tau}) =
e^{i\varphi}  \sum_{a^{\prime}=1}^{p}S_{a,a^{\prime}} 
\Theta^{ NS}_{2a^{\prime}}(\t,\z),   \\
    &  \Theta^{ \widetilde{NS}}_{2a} (-\frac{1}{\tau},
-\frac{\zeta}{\tau}) =e^{i\varphi}  \sum_{a^{\prime}=1}^{p}
S_{a,a^{\prime}} \Theta^{ R}_{2a^{\prime}}(\t,\z), \notag \\
    &\Theta^{ R}_{2a} (-\frac{1}{\tau},-\frac{\zeta}{\tau}) 
= e^{i\varphi} \sum_{a^{\prime}=1}^{p}S_{a,a^{\prime}} 
\Theta^{ \widetilde{NS}}_{2a^{\prime}}(\t,\z),  \notag \\
    & \Theta^{\widetilde{R}}_{2a} (-\frac{1}{\tau},
-\frac{\zeta}{\tau}) =- e^{i\varphi} \sum_{a^{\prime}=1}^{p}S_{a,a^{\prime}} 
\Theta^{ \widetilde{R}}_{2a^{\prime}}(\t,\z), \notag 
\end{align}
with
\be
\label{Sjain-el}
S_{a,a^{\prime}}=\frac{ 1 }{\sqrt{p}}  \text{exp} 
\bigg( 2\pi i \frac{2aa^{\prime}}{p} \bigg),
~~~~~~e^{i\varphi} =\text{exp}\bigg(  i \frac{\pi}{2p}\text{Re}\bigg(
\frac{(2\zeta)^2}{\tau} \bigg)\bigg) .
\ee
\item $T$
\begin{align}
\label{Tjain-tranf}
&\Theta^{NS}_{2a}(\tau+1, \zeta)= T_a 
\Theta^{ \widetilde{NS}}_{2a}(\tau, \zeta), \\
    &\Theta^{ \widetilde{NS}}_{2a}(\tau+1, \zeta)= 
T_a\Theta^{NS}_{2a}(\tau, \zeta),\notag\\
    &\Theta^{ R}_{2a}(\tau+1, \zeta)= 
\text{exp}\bigg( \frac{i \pi}{2} \bigg)T_a \Theta^{ R}_{2a}(\tau, \zeta), 
 \notag \\
    &\Theta^{  \widetilde{R}}_{2a}(\tau+1, \zeta)= 
\text{exp}\bigg( \frac{i \pi}{2} \bigg)
T_a \Theta^{ \widetilde{R}}_{2a}(\tau, \zeta), \notag 
\end{align}
with
\be
\label{Tjain-el}
T_a= \text{exp} \big( 2 \pi i \big( \frac{a^2}{p}-\frac{1}{12}  \big) \big).
\ee
\item $V$
\begin{align}
\label{Vjain-tranf}
&\Theta^{ NS}_{2a}(\tau, \zeta + \tau) =  
 V_{\Phi_0}  \Theta^{ NS}_{2a +2}(\tau, \zeta),  \\
    &\Theta^{ \widetilde{NS}}_{2a}(\tau, \zeta + \tau) =  
 V_{\Phi_0}  \Theta^{ \widetilde{NS}}_{2a +2}(\tau, \zeta),\notag \\
    &\Theta^{ R}_{2a}(\tau, \zeta + \tau) =  
 V_{\Phi_0}  \Theta^{ R}_{2a +2}(\tau, \zeta),\notag\\
    &\Theta^{ \widetilde{R}}_{2a}(\tau, \zeta + \tau) = 
  V_{\Phi_0}  \Theta^{\widetilde{R}}_{2a +2}(\tau, \zeta), \notag
\end{align}
with
\be
\label{Vjain-el}
V_{\Phi_0}=  \text{exp} \big( -2 \pi i \frac{2}{p}
 \big( Re \frac{ \tau}{2} +  Re \zeta \big) \big).
\ee
\item  $V^{\frac{p}{2}}$
\begin{align}
\label{Vp2jain-tranf}
&\Theta^{NS}_{2a} (\tau,\zeta + \frac{p \t}{2}) =
 V_{\frac{p}{2}\Phi_0} \Theta^{ R}_{2a}(\tau,\zeta),    \\
    & \Theta^{ \widetilde{NS}}_{2a} (\tau,\zeta + \frac{p \t}{2}) = 
V_{\frac{p}{2}\Phi_0} \Theta^{ \widetilde{R}}_{2a}(\tau,\zeta), \notag\\
    &\Theta^{ R}_{2a} (\tau,\zeta + \frac{p \t}{2}) = 
V_{\frac{p}{2}\Phi_0} \Theta^{ NS}_{2a}(\tau,\zeta),\notag\\
    & \Theta^{ \widetilde{R}}_{2a} (\tau,\zeta + \frac{p \t}{2}) = 
V_{\frac{p}{2}\Phi_0} \Theta^{ \widetilde{NS}}_{2a}(\tau,\zeta),\notag
    \end{align}
with
\be
\label{Vp2jain-el}
V_{\frac{p}{2}\Phi_0}= \text{exp} \big( -2\pi i \big( 
\frac{p}{4}Re\tau + Re\zeta \big) \big).
\ee
\end{itemize}
Using these transformations, we obtain that the partition
functions for the Jain states (\ref{jain-pf-tot}) transform
as in the $c=1$ case under the $V^{\frac{p}{2}}$ and modular
transformations, because the number of charge sectors $p$ is odd
for the Jain theory (see Fig.(\ref{fig6}) and Section 4.2.1). Owing to
the absence of $Z_2$ anomaly, the modular invariant partition function
$Z_{\text{Ising}}$ is consistent with TR symmetry, then the
topological phase is instable.

In the following, we also give the partition functions
for the general multicomponent Jain theory with $c=k$ and symmetry
$\widehat{U(1)}\times \widehat{SU(k)}_1$. The stability
parameters are $(k,p)=(k,2nk+1),~n\in\mathbb{N}$. The form of the
transformations depends on the parity of the $k$ parameter:

i) For even $k$, we remember from Ref. \cite{cz-mod} 
that the  charge characters
are given by $K_{ka+\alpha p}(\tau,k\zeta; kp)$, with
$\a=1,\cdots,k$ and $a=1,\cdots,p$. The neutral characters $\c(\t,0)$
belong to the $\widehat{SU(k)}_1$ affine algebra and have the
periodicity and modular transformations (\ref{mod-neutro-jain}).
Starting from the anyon partition function of the Neveu-Schwarz
sector (\ref{JainNS-pf}), we can act with the $T$ and $ST$ modular
transformations to obtain the other spin sectors. Altogether they are:
\begin{align}
\label{anyon-jain-even}
    & \Theta_{ka}^{NS}(\tau,\zeta;k)= \sum_{\a=1}^k
K_{ka+\a p}(\tau, k\zeta; kp) ~ \chi_{\a}(\tau,0), \\ &
\Theta_{ka}^{\widetilde{NS}}(\tau,\zeta;k)= \sum_{\a=1}^k(-1)^{\a}
K_{ka+\a p}(\tau, k\zeta; kp)~  \chi_{\a}(\tau,0),\notag\\ 
& \Theta_{ka}^{R}(\tau,\zeta;k)= \sum_{\a=1}^k
K_{ka+\a p}(\tau, k\zeta; kp) ~  \chi_{\a+\frac{k}{2}}(\tau,0), \notag \\ &
 \Theta_{ka}^{\widetilde{R}}(\tau,\zeta;k)=
\sum_{\a=1}^k (-1)^{\a+\frac{k}{2}} K_{ka+\a p}(\tau, k\zeta; kp) ~
 \chi_{\a+\frac{k}{2}}(\tau,0).\notag
\end{align}
Note that for $k=2$ we obtain the expressions given before
(Eq.(\ref{jain-anyon2})).

ii) For odd $k$, the charge characters are different for each spin
sector (as in the $c=1$ case). For the $NS$ and $\wt{NS}$ spin
sectors they read: 
\be
\label{jain-kodd-NS}
K^{NS}_{ka+\alpha p}(\tau,k\zeta; kp)=K_{ka+\alpha p}(\tau,k\zeta; kp),
\ee
\be
\begin{split}
&K^{\widetilde{NS}}_{ka+\alpha p}(\tau,k\zeta; kp)= \\
&\frac{F(\t,\z)}{{\eta(\tau)} }
\sum_{n\in\mathbb{Z }} (-1)^{pkn}\text{ exp} 
\bigg[ 2\pi i \bigg( \frac{\tau}{2kp}
  \bigg(pkn+ka+\alpha p \bigg)^2 +
  \frac{\zeta}{p}\bigg(pkn+2a+\alpha p\bigg)+
\frac{ka}{2}+\frac{\a}{2} \bigg) \bigg].\notag
  \end{split}
\ee
The corresponding expressions for the $R$ and $\wt{R}$ spin sectors
are defined by
\begin{align}
\label{jain-kodd-R}
    &K^{R}_{ka+\alpha p}= K^{NS}_{ka+\alpha p+\frac{kp}{2}}, \\ &
K^{\widetilde{R}}_{ka+\alpha p}=
K^{\widetilde{NS}}_{ka+\alpha p+\frac{kp}{2}}\notag.
\end{align}
Upon combining them with the neutral characters $\c_{\a}(\t,0)$ of the
$\widehat{SU(k)}_1$ affine algebra, the anyon sectors take the general
form for the simple current modular invariants, as explained in Section 4: 
\be
\label{any-kodd-jain}
\Theta_{ka}^{(\s)}(\t,\z;k)=\sum_{\a=1}^k K^{(\s)}_{ka+\a
  p}(\t,k\z;kp) \c_{\a}(\t,0),~~~~\s=NS, \widetilde{NS}, R,
\widetilde{R}.  
\ee 
The transformations under the modular group and
the insertion fluxes are those shown in Fig.{\ref{fig6}}, since $p$ is
always odd in the Jain theory.

\subsection{General $U(1)^n$ Abelian states}
\label{app.cn}
Let us discuss the general Abelian states whose spectrum of charge
and statistics is described by the $K$-matrix formalism (see Section
4.3.2). In the case that the $n$ chiral
(spin-up) and $n$ antichiral (spin-down) edge modes are independent,
the ${\bf K}$ matrix takes the block
diagonal form (\ref{block-K}) and the $n\times n$ block 
$K$ is definite positive, integer valued and with $K_{ii}$ odd, 
$i=1,\cdots, n$.  In the basis in which the charge vector is
${\bf t}=(t,t)$, where $t=(1,\cdots,1)$ $n-$dimensional,  
the characters of the chiral modes have been written in 
Ref.\cite{cz-mod}. The chiral anyon sectors modes of the $NS$ and
$\wt{NS}$ spin sectors are (with 
$\boldsymbol{\lambda} \in \mathbb{Z}^n/K\mathbb{Z }^n $):
\begin{equation}
\label{NS-abelian}
K^{NS}_{\boldsymbol{\lambda}}(\tau,\zeta) = 
\frac{F(\t,\z)}{\eta({q})^n}\sum_{\ell \in\mathbb{Z }^n} \text{exp}
  \biggl( 2\pi i \biggl[ \frac{\tau}{2}(K\ell+\lambda)^T K^{-1}
    (K\ell+\lambda) +\zeta t^T (\ell +K^{-1}\lambda) \biggr]\biggr),
\end{equation}
\begin{equation}
\begin{split}
&K^{\widetilde{NS} }_{\boldsymbol{\lambda}}(\tau,\zeta)=
  \frac{F(\t,\z)}{\eta(q)^n} \times \\
& \sum_{\ell\in\mathbb{Z }^n} (-1)^{\ell^T b}
  \text{exp} \biggl( 2\pi i \biggl[ \frac{\tau}{2}(K\ell+\lambda)^T K^{-1}
    (K\ell+\lambda) +\zeta t^T (\ell +K^{-1}\lambda) +\frac{1}{2}\lambda^T
    K^{-1}b \biggr]\biggr),\notag
\end{split}
\end{equation} 
with the prefactor $F(\t,\z)=\text{exp} \biggl(-\pi
  t^{T}K^{-1}t\, (\I\zeta)^2/\I \tau \biggr)$; 
those of the $R$ and $\wt{R}$ sectors are related to these by
\begin{align}
\label{R-abelian}
&K^{R}_{\boldsymbol{\lambda}}= K^{
  NS}_{\boldsymbol{\lambda}+\frac{\mathbf{ b}}{2}},\\ &K^{
  \widetilde{R} }_{\boldsymbol{\lambda}}= K^{\widetilde{ NS}
}_{\boldsymbol{\lambda}+\frac{\mathbf{ b}}{2}}.\notag
\end{align}
In these equations, $\textbf{b}$ is a $n$-dimensional vector whose
components are the diagonal elements of the $K$ matrix:
\begin{equation}
\label{b-vector}
\mathbf{b}=\begin{pmatrix} K_{11}\\ \vdots \\ K_{nn} \end{pmatrix}.
\end{equation}
The edge partition functions for each spin sectors
are
\begin{equation}
\label{pf-abelian}
Z^{(\s)}(\tau,\zeta)=\sum_{{\boldsymbol{\lambda}} \in \mathbb{ Z}^n /K\mathbb{
    Z}^n} K^{(\s)}_{\boldsymbol{\lambda}}(\tau,\zeta)\ov{{K}^{
  (\s)}_{\boldsymbol{-\lambda}}}(\tau,\zeta),~~~~~ \s=NS, \widetilde{NS}, R,
\widetilde{R}.
\end{equation}

As in previous cases, we give the transformation rules of the anyon
sectors (\ref{NS-abelian}) and (\ref{R-abelian}):
\begin{itemize}
\item $S$
 \begin{align}
\label{S-abel}
    & K^{ NS}_{\boldsymbol{\lambda}}(-\frac{1}{\tau},\frac{-\zeta}{\tau}) = 
\text{e}^{i\varphi}\sum_{{\boldsymbol{\lambda}^{\prime}} 
\in \mathbb{ Z}^n /K\mathbb{ Z}^n} 
S_{\boldsymbol{\lambda},\boldsymbol{\lambda}^{\prime}} 
~K^{NS}_{\boldsymbol{\lambda}^{\prime}}(\tau,\zeta),  \\
    &  K^{\widetilde{ NS} }_{\boldsymbol{\lambda}}(-\frac{1}{\tau},
\frac{-\zeta}{\tau}) =\text{exp}\bigg( 2 \pi i ~\lambda^T K^{-1} b \bigg) 
~\text{e}^{i\varphi}\sum_{{\boldsymbol{\lambda}^{\prime}}
 \in \mathbb{ Z}^n /K\mathbb{ Z}^n} S_{\boldsymbol{\lambda},
\boldsymbol{\lambda}^{\prime}}
K^{ R}_{\boldsymbol{\lambda}^{\prime}}(\tau,\zeta),\notag\\
    &K^{R}_{\boldsymbol{\lambda}} (-\frac{1}{\tau},\frac{-\zeta}{\tau}) = 
\text{e}^{i\varphi}\sum_{{\boldsymbol{\lambda}^{\prime}} 
\in \mathbb{ Z}^n /K\mathbb{ Z}^n} S_{\boldsymbol{\lambda},
\boldsymbol{\lambda}^{\prime}} 
K^{\widetilde{NS} }_{\boldsymbol{\lambda}^{\prime}}(\tau,\zeta),\notag   \\
    &K^{\widetilde{R} }_{\boldsymbol{\lambda}} 
(-\frac{1}{\tau},\frac{-\zeta}{\tau}) =  
\text{exp}\bigg( 2 \pi i \bigg(\lambda^T K^{-1} b 
+\frac{1}{4}b^T K^{-1}b\bigg)\bigg)~   \text{e}^{i\varphi}
\sum_{{\boldsymbol{\lambda}^{\prime}} \in \mathbb{ Z}^n /K\mathbb{ Z}^n} 
S_{\boldsymbol{\lambda},\boldsymbol{\lambda}^{\prime}}
K^{\widetilde{ R} }_{\boldsymbol{\lambda}^{\prime}}(\tau,\zeta)\notag,
\end{align}
with 
  \begin{equation}
  \label{S-abel-el}
  S_{\boldsymbol{\lambda},\boldsymbol{\lambda}^{\prime}}(\tau, \zeta)= 
\frac{  \text{exp}\bigg(\bigg( 2 \pi i \lambda^T K^{-1}
\lambda^{\prime} \bigg)\bigg)  }{ \sqrt{\text{det}K}   },
~~~~~\text{e}^{i\varphi}=\text{exp} \bigg( i \pi t^TK^{-1}t 
Re\bigg( \frac{\zeta^2}{\tau} \bigg)   \bigg) 
  \end{equation}
\item $T$
\begin{align}
\label{T-abel}
    & K^{ NS}_{\boldsymbol{\lambda}} (\tau+1,\zeta) = 
\text{exp}\bigg( -2 \pi i~ \frac{1}{2}\lambda^T K^{-1}b \bigg)
 T_a ~K^{\widetilde{ NS} }_{\boldsymbol{\lambda}}(\tau,\zeta),  \\
    &   K^{\widetilde{ NS} }_{\boldsymbol{\lambda}}(\tau+1,\zeta) =
 \text{exp}\bigg( 2 \pi i~ \frac{1}{2}\lambda^TK^{-1}b \bigg) 
T_a~ K^{ NS }_{\boldsymbol{\lambda}}(\tau,\zeta),\notag   \\
    &K^{R}_{\boldsymbol{\lambda}}(\tau+1,\zeta) = 
T_a~T_b  ~K^{ R}_{\boldsymbol{\lambda}}(\tau,\zeta),\notag\\
    &K^{\widetilde{ R} }_{\boldsymbol{\lambda}} (\tau+1,\zeta) = 
T_aT_b~K^{\widetilde{ R} }_{\boldsymbol{\lambda}}(\tau,\zeta),\notag
\end{align}
with
\begin{equation}
\label{T-abel-el}
T_a=\text{exp}\bigg(2\pi i\bigg( \frac{1}{2}\lambda^T K^{-1} \lambda -
\frac{n}{24} \bigg) \bigg),
~~T_b=\text{exp}\bigg(2\pi i\bigg( \frac{1}{2}\lambda^T K^{-1} b +
\frac{1}{8}b^T K^{-1}b \bigg) \bigg).
\end{equation}
In order to obtain these transformations, we used 
the symmetry of the $K$ matrix and
the odd parity of its diagonal elements, leading to:
\begin{equation}
\label{parity-kmatrix}
\text{exp}\bigg[2 \pi i \frac{1}{2}\ell^T K \ell\bigg] =
 \text{exp}\bigg[ 2 \pi i \frac{1}{2} K_{ii}\ell^2_i \bigg]= (-1)^{\ell^T b}.
\end{equation}
\item $V$
 \begin{align}
\label{V-abel}
    &K^{ NS}_{\boldsymbol{\lambda}} (\tau,\zeta+ \tau) =
V_{\Phi_0}K^{ NS }_{\boldsymbol{\lambda} +\mathbf{ t}}(\tau,\zeta), \\
    &  K^{\widetilde{ NS}}_{\boldsymbol{\lambda}} (\tau,\zeta+\tau) =
\text{exp}\bigg(-2 \pi i ~\frac{1}{2}t^TK^{-1}b \bigg)
~V_{\Phi_0}K^{\widetilde{ NS} }_{\boldsymbol{\lambda}+\mathbf{ t}}(\tau,\zeta),
\notag   \\
    &K^{ R}_{\boldsymbol{\lambda}} (\tau,\zeta+\tau) =
V_{\Phi_0}K^{ R }_{\boldsymbol{\lambda}+\mathbf{ t}}(\tau,\zeta),\notag   \\
    &K^{\widetilde{ R}}_{\boldsymbol{\lambda}} (\tau,\zeta+\tau) =
\text{exp}\bigg(-2 \pi i ~\frac{1}{2}t^TK^{-1}b \bigg)~V_{\Phi_0} 
K^{\widetilde{R} }_{\boldsymbol{\lambda}+\mathbf{ t}}(\tau,\zeta),\notag
\end{align}
with
\be
\label{V-label-el}
V_{\Phi_0}(\tau, \zeta)= \text{exp} \bigg( -2 \pi i t^T K^{-1}t
 \bigg( Re\frac{\tau}{2} +Re \zeta \bigg)  \bigg).
\ee
\end{itemize}

In this theory, the partition functions (\ref{pf-abelian}) transform 
under the modular group as depicted
in Fig.(\ref{fig6})(b).  We have to more careful about the
$V^{\frac{p}{2}}$ transformation: according to the discussion 
in Section (4.2.1), there are two different
results depending on the parity of $p$. This number can be obtained from
the minimal value $e^*=1/p$ in the charge spectrum.
In general, we can write:
\begin{align}
\label{Vp2-abel}
    &  K^{ NS}_{\boldsymbol{\lambda}} (\tau,\zeta+\frac{p \t}{2})= 
V_{\frac{p}{2}}~K^{ NS}_{\boldsymbol{\lambda}+\frac{p}{2}\textbf{t}}(\tau,\zeta), \\
    &  K^{ \widetilde{NS}}_{\boldsymbol{\lambda}} (\tau,\zeta+\frac{p \t}{2})=
 V_{\frac{p}{2}}~\text{exp}\bigg( -2\pi i~\frac{p}{4}t^T K^{-1}b \bigg)
~K^{  \widetilde{NS}}_{\boldsymbol{\lambda}+\frac{p}{2}\textbf{t}}(\tau,\zeta),
\notag \\
    &K^{ R}_{\boldsymbol{\lambda}} (\tau,\zeta+\frac{p \t}{2})= 
V_{\frac{p}{2}}~K^{ R}_{\boldsymbol{\lambda}+\frac{p}{2}\textbf{t}}(\tau,\zeta), 
\notag\\
    &K^{ \widetilde{R}}_{\boldsymbol{\lambda}} (\tau,\zeta+\frac{p \t}{2})= 
V_{\frac{p}{2}}~\text{exp}\bigg( -2\pi i~\frac{p}{4}t^T K^{-1}b \bigg)
~K^{  \widetilde{R}}_{\boldsymbol{\lambda}+\frac{p}{2}\textbf{t}}(\tau,\zeta),\notag 
\end{align}
with
\begin{equation}
\label{Vp2-abel-el}
V_{\frac{p}{2}}= \text{exp}\bigg(-2 \pi i~ t^T K^{-1}t
 \bigg( \frac{p^2}{8}Re\tau +\frac{p}{2}Re \zeta \bigg) \bigg).
\end{equation}
We distinguish the two cases:

i) If $p$ is odd, the $V^{\frac{p}{2}}$ transformation maps
$NS$ sectors into $R$ sectors, up to a reshuffling of the anyon label, 
$\l\to \l'$, as follows:
\be
\label{Vp2resch}
V^{\frac{p}{2}}: K^{NS}_{\boldsymbol{\lambda}} \to
K^{R}_{\boldsymbol{\lambda}'}, 
\ee 
and similarly for the other
sectors. Then the partition function of the Neveu-Schwarz sector is
mapped in that of the Ramond sector and so on, as depicted in
Fig.(\ref{fig6})(a).

ii) If $p$ is even, every spin sector partition function returns to
itself, up to a reshuffling of anyon sectors in
Eq.(\ref{pf-abelian}), 
\begin{equation}
\label{Vp2-easy}
V^{\frac{p}{2}}: Z^{\s}(\tau,\zeta) \to
Z^{\s}(\tau,\zeta+\frac{p \t}{2})=Z^{\s}(\tau,\zeta) ,~~~~~\s=NS,
\widetilde{NS}, R, \widetilde{R},
\end{equation}
as is pictorially represented in Fig.(\ref{fig7})(a).

\subsection{(331) and Pfaffian states}
\label{app.331}
For the (331) state, we recall from the main text that 
$(k,p)=(2,4)$, the charge character is $K_{\lambda}(\tau, 2\zeta;8)$,
with $\l=1,\cdots, 8$, and the Abelian neutral character is
$\chi_{\lambda}(\tau,4)=K_{\lambda}(\tau, 0; 4)$ \cite{cgt1}.  
The anyon sectors for the chiral (spin-up) modes are:
\begin{align}
\label{331-any-spin}
    &\Theta_{\lambda}^{NS} (\tau,\zeta; 2) =
K_{\lambda}~\chi_{\lambda} +
K_{\lambda+4}~\chi_{\lambda+2}\\ &\Theta_{\lambda}^{\widetilde{NS}}
(\tau,\zeta; 2) = K_{\lambda}~\chi_{\lambda} -
K_{\lambda+4}~\chi_{\lambda+2}\notag\\ &\Theta_{\lambda}^{R}
(\tau,\zeta; 2) = K_{\lambda}~\chi_{\lambda+1} +
K_{\lambda+4}~\chi_{\lambda+3}\notag\\ &\Theta_{\lambda}^{
  \widetilde{R}} (\tau,\zeta; 2) = K_{\lambda}~\chi_{\lambda+1} -
K_{\lambda+4}~\chi_{\lambda+3}.\notag
   \end{align}
Then, the edge partition functions for the Neveu-Schwarz and Ramond
spin sectors of the (331) topological states can be written as:
\begin{align}
\label{331-pf}
    &Z^{NS/ \widetilde{NS}}_{(331)}= \sum_{\lambda=0}^3 \bigg\{ \big|
K_{\lambda} ~\chi_{\lambda} ~\pm~K_{\lambda+4}~\chi_{\lambda+2} \big|^2 +
\big| K_{\lambda} ~\chi_{\lambda+2}~\pm~K_{\lambda+4}~\chi_{\lambda} \big|^2
\bigg\}, \\ & Z^{R/\widetilde{R}}_{(331)}= \sum_{\lambda=0}^3 \bigg\{ \big|
K_{\lambda} ~\chi_{\lambda +1}\pm K_{\lambda+4}~\chi_{\lambda+3} \big|^2 +
\big| K_{\lambda} ~\chi_{\lambda+3} \pm K_{\lambda+4}~\chi_{\lambda+1} \big|^2
\bigg\}.\notag
\end{align}
The (331) anyon sectors (\ref{331-any-spin}) transform under the
modular group and the insertion of  fluxes according to the Abelian formulas
Eq.(\ref{S-abel}-\ref{Vp2-abel}) with $n=2$, and $p=4$ even. 

The edge partition function for the Pfaffian states in the
Neveu-Schwarz spin sector is given in (\ref{pfaffian-NS})
\cite{cv}. We obtain the partition functions of the other spin sectors
acting with $T$ and $ST$ on $Z^{NS}_{\text{Pf}}$ (\ref{pfaffian-NS})
as described in Section (3.2). They read:
\begin{equation}
\begin{split}
\label{Pfaffian-pf}
     Z^{NS/ \widetilde{NS}}_{\text{Pf}} =& \big| K_0 I \pm K_4 \psi
     \big|^2 + \big| K_0\psi \pm K_4 I \big|^2 + \big| (K_1\pm K_{-3})\sigma
     \big|^2 \\ & \big| K_2I \pm K_{-2}\psi \big|^2 + \big| K_2\psi\pm K_{-2}I
     \big|^2 + \big| (K_3+K_{-1})\sigma \big|^2,
    \end{split}
\end{equation}
\begin{equation}
\begin{split}
     Z^{R/ \widetilde{R}}_{\text{Pf}} =& \big| K_3 I \pm K_{-1} \psi
     \big|^2 + \big| K_3\psi \pm K_{-1} I \big|^2 + \big| (K_0\pm K_{4})\sigma
     \big|^2 \\ & \big| K_{-3}I \pm K_{1}\psi \big|^2 + \big| K_{-3}\psi\pm
     K_{1}I \big|^2 + \big| (K_2+K_{-2})\sigma \big|^2.\notag
    \end{split}
\end{equation}
We have written the neutral characters in (\ref{Pfaffian-pf}) with the
same symbol of the Ising fields
\begin{equation}
\label{ising field}
\chi_0^0=\chi_2^2=I,~~~~~\chi_1^1=\chi_3^1=\sigma,~~~~~\chi^0_2=\chi^2_0=\psi,
\end{equation}
that model the neutral excitations of this system \cite{cft} \cite{mr}. 

The spin sectors partition functions of (331) and Pfaffian topological
states, Eqs. (\ref{331-pf}) and (\ref{Pfaffian-pf}), respectively,
transform in the same manner under $V^{\frac{p}{2}}$ and the modular
group because the two theories have the same $(k,p)$ parameters
entering the stability analysis, as discussed in Section
(4.3.3). Since $p$ is even, these transformations are shown in
Fig.(\ref{fig7}). The absence of the $Z_2$ anomaly in both cases
allows for the $Z_{\text{Ising}}$ modular invariant (\ref{Ising}) to
be TR symmetric, such that the two topological phases are unstable.

\subsection{Read-Rezayi states}
\label{app.rr}
The Read-Rezayi models \cite{rr} are based on the neutral
$\mathbb{Z}_k$ parafermion conformal theories with central charge
$c=2(k-1)/(k+2)$, described by the coset construction
$\wh{SU(2)_k}/\wh{U(1)_{2k}}$ \cite{cv}.  For these theories the values
of the stability parameters are $p=kM+2$ with $M=1,3,5,\cdots$ and
$k=2,3,\cdots$. The structure of partition functions depends on the
parity of $k$.

\subsubsection{Partition functions for RR states with even $k$}
\label{RRkeven_pf}
The charge characters are given by the functions (\ref{pf})
$K_{\lambda}(\tau, k\zeta, kp)$ with periodicities
$K_{\l+kp}=K_{\l}$. The $\mathbb{Z}_k$ parafermionic characters that
describe the neutral part are denoted by $\chi^{\ell}_m(\tau;2k)$, and
have the following periodicities and modular transformations
\cite{cv},
  \begin{align}
\label{para-period}
    &\chi^{\ell}_m=\chi^{\ell}_{m+2k}=\chi^{k-\ell}_{m+k},
~~~~m=\ell~\text{mod}~2,
\\ & \chi^{\ell}_m=0~~~~~~~~~~~~~~~~~~~~~~~m=\ell+1~\text{mod}~2\notag,\\
\label{mod-neutro-rr}
    &S:~~~~~\chi_{m}^{\ell}(-\frac{1}{\tau},0; 2k) = 
\frac{1}{\sqrt{2k}}\sum_{\ell^{\prime}=0}^{k} 
\sum_{m^{\prime}=1}^{2k}\text{exp}\bigg( -2\pi i \frac{mm^{\prime}}{2k} \bigg)  
 s_{\ell,\ell'}~\chi_{m^{\prime}}^{\ell^{\prime}}(\tau; 2k),\\
    & ~~~~~~~~~~s_{\ell,\ell'}=
\sqrt{\frac{2}{k+2}}~\text{sin}\bigg( \frac{\p(\ell+1)(\ell'+1)}{k+2}\bigg),
\notag\\
    & T:~~~~~ \chi_{m}^{\ell}(\tau+1,0; 2k)=
\text{exp}\bigg(2\pi i\bigg( \frac{\ell(\ell+2)}{4(k+2)} -
\frac{m^2}{4k}+\frac{1}{24} \bigg)\bigg) \chi_{m}^{\ell}(\tau; 2k).\notag
\end{align}
Starting from the Neveu-Schwarz anyon sectors given in Ref.\cite{cv}
and acting with $T$ and $ST$, we find those of the other spin
sectors. Altogether they read:
\begin{align}
\label{rr-anyon}
    & \Theta_a^{\ell  NS}(\tau,\zeta;k)= \sum_{b=1}^k
K_{a+bp}(\tau, k\zeta; kp) ~ \chi_{a+2b}^{\ell}(\tau; 2k), \\ &
\Theta_a^{\ell  \widetilde{NS}}(\tau,\zeta;k)= \sum_{b=1}^k(-1)^b
K_{a+bp}(\tau, k\zeta; kp)~ \chi_{a+2b}^{\ell}(\tau;
2k),\notag\\ & \Theta_a^{\ell  R}(\tau,\zeta;k)= \sum_{b=1}^k
K_{a+bp}(\tau, k\zeta; kp) ~ \chi_{a+2b+\frac{k}{2}}^{\ell}(\tau;
2k),\notag \\ & \Theta_a^{\ell  \widetilde{R}}(\tau,\zeta;k)=
\sum_{b=1}^k (-1)^b K_{a+bp}(\tau, k\zeta; kp) ~
\chi_{a+2b+\frac{k}{2}}^{\ell}(\tau; 2k),\notag
\end{align}
where $a=0,1,\cdots, p-1$, and $\ell=0,1,\cdots,k$.  
The partition functions in the corresponding spin sectors are:
\begin{equation}
\label{rr-keven-pf}
Z^{(\s)}_{RR}= \sum_{\ell=0}^{k} \sum_{\substack{a=1\\a=\ell
    ~\text{mod}~2}}^{p} \Theta_a^{\ell  (\s)}
~\ov{\Theta_{-a}^{\ell  (\s)}} , ~~~~~~~\s=NS, \widetilde{NS}, R,
\widetilde{R}.
\end{equation}
We note that the partition functions of the Pfaffian state
(Eq.\ref{Pfaffian-pf}) are obtained by choosing $M=1$ and $k=2$ in the
previous formulas. The transformations of the anyon sectors
(\ref{rr-anyon}) and the partition functions (\ref{rr-keven-pf}) under
the insertion of fluxes and the modular group are the same of
the Pfaffian state, represented in Fig.(\ref{fig7}).  In this case the
$Z_2$ anomaly is absent, the modular invariant partition function
$Z_{\text{Ising}}$ (\ref{Ising}) is consistent with TR symmetry and
the topological phases are unstable.

\subsubsection{Partition functions for RR states with odd $k$}
\label{RRkodd_pf}
It is necessary to introduce different charge characters for
each spin sector. For the $NS$ and $\wt{NS}$ sectors they are:
\begin{align}
\label{rr-kodd-NS}
    & K^{NS}_{\lambda}(\tau, k\zeta;
kp)= K_{\lambda}(\tau, k\zeta;
kp),\\ &
K^{\widetilde{NS}}_{\lambda}(\tau, k\zeta;
kp)=\frac{F(\t,\z)}{{\eta(\tau)} }\sum_{n\in\mathbb{Z}} 
(-1)^{nkp} \text{exp} {\bigg[2\pi
    i \bigg( \frac{\tau}{2kp}\big(nkp+\lambda \big)^2 +
    \frac{\zeta}{p}\big(nkp+ \lambda \big) +\frac{\lambda}{2}
    \bigg)\bigg]}.\notag
    \end{align}
Those of the $R$ and $\wt{R}$ spin sectors are defined by:
\begin{align}
\label{rr-kodd-R}
    &K^{R}_{\lambda}= K^{NS}_{\lambda+\frac{kp}{2}}, \\ &
K^{\widetilde{R}}_{\lambda}=
K^{\widetilde{NS}}_{\lambda+\frac{kp}{2}}\notag.
\end{align}
These charge characters have the same periodicities (\ref{pk-period}),
i.e. $K_{\l}=K_{\l+kp}$.  Making use of $\mathbb{Z}_k$ parafermionic
characters $\chi^{\ell}_m(\tau;2k)$, the anyon sectors for 
four spin sector take the usual form of simple current invariants:
\begin{align}
\label{rr-kodd-anyon}
    & \Theta_a^{\ell (\s)}(\tau,\zeta;k)= \sum_{b=1}^k
K^{(\s)}_{a+bp}(\tau, k\zeta; kp) ~ \chi_{a+2b}^{\ell}(\tau;
2k),~~~~\s= NS, \widetilde{NS}, R, \widetilde{R},
\end{align}
The edge partition functions of the four spin sector have the same
expression of those in the even $k$ case \eqref{rr-keven-pf}.  Because
the values of $(k,p)$ are odd, as discussed in Section (4.2.1), it is
easy to show that under the $V^{\frac{p}{2}}$ transformation and
the modular group the anyon sectors and the partition functions
transform as  represented in Fig.(\ref{fig6}). Owing
to the presence of $Z_{2}$ anomaly, the modular invariant partition
function $Z_{\text{Ising}}$ is inconsistent with TR symmetry, and
these non-Abelian topological phases are stable.


\end{document}